\documentclass{article}

\usepackage{microtype}
\usepackage{graphicx}
\usepackage{multirow}
\usepackage{subfigure}
\usepackage{enumitem}
\usepackage{booktabs} % for professional tables
\usepackage{amssymb}
\usepackage{amsmath}
\usepackage[font=small,labelfont=bf]{caption}

\usepackage[T1,hyphens]{url}

\usepackage{hyperref}

\newcommand{\beginsupplement}{%
        \setcounter{table}{0}
        \renewcommand{\thetable}{S\arabic{table}}%
        \setcounter{figure}{0}
        \renewcommand{\thefigure}{S\arabic{figure}}%
        \setcounter{section}{0}
        \renewcommand{\thesection}{S\arabic{section}}%
     }

\usepackage[accepted]{icml2021}

\icmltitlerunning{Gradient Disaggregation: Breaking Privacy in Federated Learning by Reconstructing the User Participant Matrix}

\begin{document}

\twocolumn[
\icmltitle{Gradient Disaggregation: Breaking Privacy in Federated Learning by Reconstructing the User Participant Matrix}



\icmlsetsymbol{equal}{*}

\begin{icmlauthorlist}
\icmlauthor{Maximilian Lam}{hv}
\icmlauthor{Gu-Yeon Wei}{hv}
\icmlauthor{David Brooks}{hv}
\icmlauthor{Vijay Janapa Reddi}{hv}
\icmlauthor{Michael Mitzenmacher}{hv} 
\end{icmlauthorlist}

\icmlaffiliation{hv}{Harvard University, Cambridge, MA}
\icmlcorrespondingauthor{Maximilian Lam}{maxlam@g.harvard.edu}

\icmlkeywords{Machine Learning, Federated Learning, Privacy, Security}

\vskip 0.3in
]



\printAffiliationsAndNotice{}  % leave blank if no need to mention equal contribution

\begin{abstract}
We show that aggregated model updates in federated learning may be insecure. An untrusted central server may disaggregate user updates from sums of updates across participants given repeated  observations, enabling the server to recover privileged information about individual users' private training data via traditional gradient inference attacks. Our method revolves around reconstructing participant information (e.g: which rounds of training users participated in) from aggregated model updates by leveraging summary information from device analytics commonly used to monitor, debug, and manage federated learning systems. Our attack is parallelizable and we successfully disaggregate user updates on settings with up to thousands of participants. We quantitatively and qualitatively demonstrate significant improvements in the capability of various inference attacks on the disaggregated updates. Our attack enables the attribution of learned properties to individual users, violating anonymity, and shows that a determined central server may undermine the secure aggregation protocol to break individual users' data privacy in federated learning.
\end{abstract}

% TODO
% - Experiment ideas
% - How to get to 10,000 users? (Appendix) 
%   - Constraint granularity
%   - Full grid search (Test the limits)
% - Count constraint noisy (+/- n, start with smaller noise)
% - Dropping count information (missing information, e.g: rounds 5-10, etc),percentage, number of rounds required
% - Expectations on number of participations probability distribution (rather than exact number)

\section{Introduction}

Federated learning is a method for collaboratively learning a shared model across multiple participants and enhances privacy by limiting data sharing \cite{fed_goog, fed_mobile, fed_system_design, kon2017federated,kon2015federated}. Participants' data privacy is preserved by sending model updates rather than raw data, which limits the amount of information that is exposed to the central server. In the context of applications, federated learning participants are edge devices such as users' smart phones or wearables, and maintaining the integrity of their data is a critical issue. Already, federated learning has been deployed by many major companies in various privacy sensitive applications including sentiment learning, next word prediction, health monitoring, content suggestion, and item ranking \cite{fed_mobile,Li_2020, fed_system_design}. Guaranteeing data privacy in these scenarios is becoming increasingly important as the topic of privacy becomes more heavily scrutinized by the greater public and by government regulations \cite{fed_goog, gov_priv_1, gov_priv_2}.

Recent research has shown that model updates may unintentionally leak information about their respective training examples \cite{geiping2020inverting,melis2018exploiting,zhu2019deep}. A central server that obtains participants' model updates may perform inference attacks to learn significant information about participants' training data, violating the core privacy principles of the federated learning paradigm. To address this critical privacy flaw, researchers have introduced methods leveraging secure multiparty computation to limit the central server's visibility into individual participants' model updates. Notably, secure aggregation \cite{secure_agg, so2020turboaggregate} has emerged as a standard security protocol which ensures that the central server may see only the final sum of model updates, rather than any individual update by itself. Thus, information learned from the aggregated model update may not be attributed to a specific user, which offers a layer of privacy against the central server. Additionally, by aggregating updates over tens to hundreds or thousands of users, updates are obfuscated to a point where most inference attacks are rendered ineffectual \cite{melis2018exploiting, geiping2020inverting, zhu2019deep}.

\begin{figure}[t]
    \centering
    \includegraphics[width=.65\linewidth]{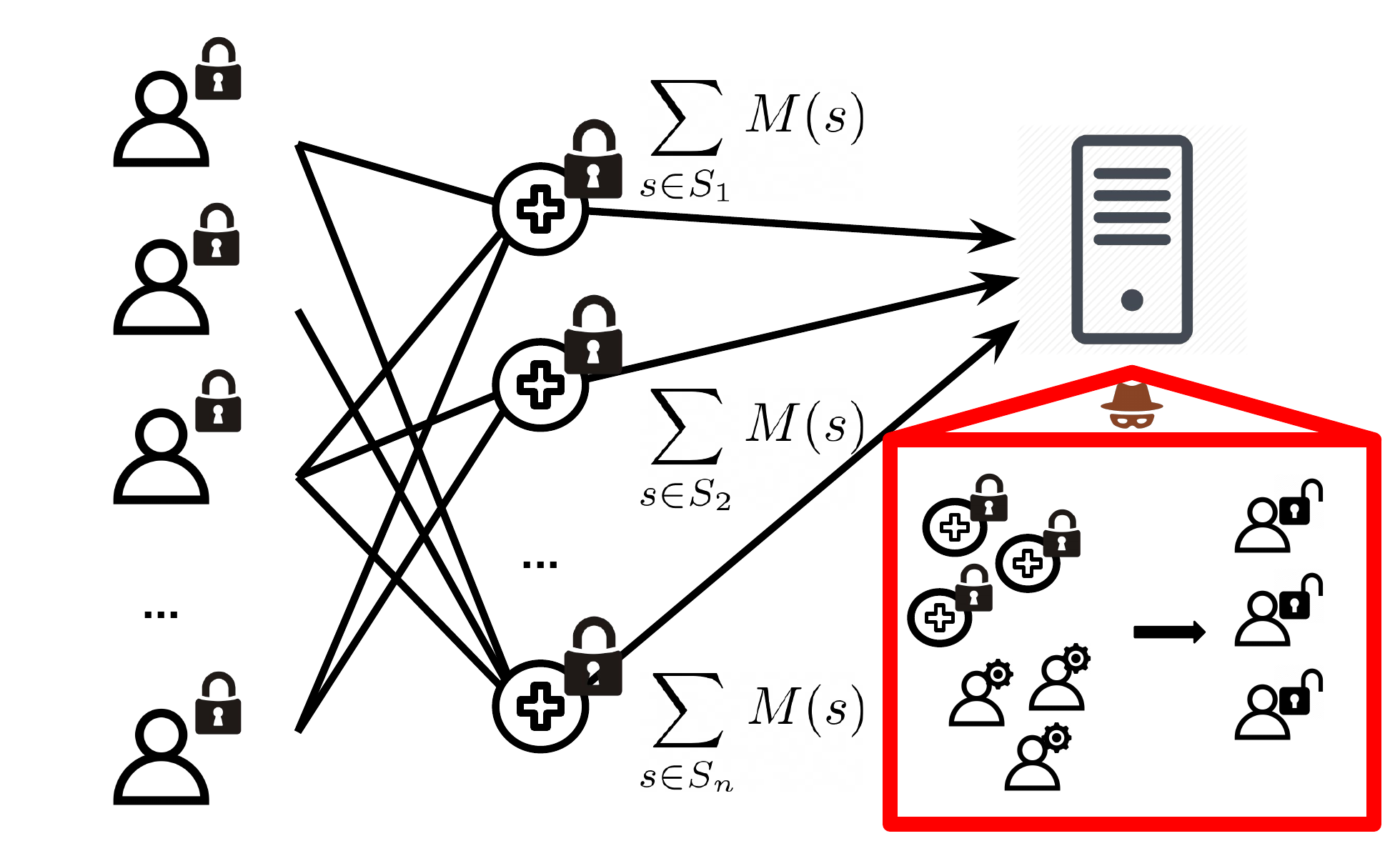}
    \caption{Our gradient disaggregation attack observes multiple rounds of aggregated model updates and leverages side channel information in the form of summary analytics collected by federated learning systems (how often users participated across certain training rounds) to uncover individual users' private model updates, undermining the secure aggregation protocol. Code: \url{https://github.com/gdisag/gradient\_disaggregation}.}
    \label{fig:grad_disaggregate}
    \vspace{-0pt}
\end{figure}

The secure aggregation protocol is secure only to the degree that it hides individual participants' model updates. A procedure that disaggregates individual participants' updates or gradients from their sum would undermine the secure aggregation protocol and unveil the aforementioned privacy vulnerability. In this work, we develop a method for gradient disaggregation, showing that secure aggregation offers little privacy protection against an adversarial server seeking to undermine individual users' data privacy. Our key insight is that participant information (e.g: which rounds of training users participated in) is derivable from aggregated model updates, when observing multiple rounds of training and leveraging summary analytics. We can reconstruct this information and use it to recover participants' individual model updates (see Figure \ref{fig:grad_disaggregate}). Our contributions are as follows:

\begin{itemize}
    \item We introduce and formulate the gradient disaggregation problem as a constrained binary matrix factorization problem. Leveraging summary analytics collected by federated learning systems, we demonstrate that our disaggregation attack can exactly recover the user participant matrix on up to thousands of participants, revealing the model update of each user. Additionally, we show that gradient disaggregation works even in the presence of significant noise and allows us to disaggregate aggregated model updates that were generated by federated averaging.
    \item We leverage gradient disaggregation to significantly improve the quality of traditional inference attacks on model updates. We show that without gradient disaggregation, inference attacks often fail to recover meaningful information on updates aggregated across tens to hundreds of users; with gradient disaggregation, we show successful recovery of users' privileged data from their disaggregated model updates. 
\end{itemize}

\section{Related Work}

\subsection{Secure Aggregation}

Secure aggregation is a method based on secure multiparty computation and is a key privacy measure deployed in federated learning systems. Secure aggregation ensures that the central server sees only the final aggregate of model updates across users while guaranteeing that no participants' updates are revealed in the clear \cite{secure_agg}. The secure aggregation protocol enhances privacy by obfuscating a user's model update with many other users' updates, limiting inference attacks such as those in \cite{melis2018exploiting, geiping2020inverting,zhu2019deep}. This obfuscation also ensures that information learned from the aggregated model update may not be attributed to an individual user. Concretely, on the issue of attribution, the secure aggregation paper states: "Using a Secure Aggregation protocol to compute these weighted averages would ensure that the server may learn only that one or more users in this randomly selected subset wrote a given word, but not which users" \cite{secure_agg}. 

% {\bf MM:  again, don't like "inherently" as the adjective; maybe "is derivable from the aggregated model updates with minimal side information"}
Our work on gradient disaggregation undermines the secure aggregation protocol by showing that, through observing multiple rounds of collected data and leveraging side channel information (specifically, user participation frequency as collected by federated learning systems), individual updates may be reconstructed from their overall sums. While secure aggregation has been proven to be cryptographically secure, leaking no information which is not leaked by the aggregated model update itself \cite{secure_agg}, the key insight of our attack is that participant information (e.g: which rounds each user participated in) is derivable from the aggregated model updates and reconstructing it allows us to in turn recover individual model updates. 

\subsection{Analytics in Federated Learning Systems}

Infrastructure to support, debug, and manage federated learning systems is critical to their functioning.  \cite{fed_system_design} outlines the design of Google's federated learning systems and describes its core components and protocols. A key aspect of their infrastructure is the collection of device analytics. Notably, \cite{fed_system_design} collect several important device metrics such as how often devices performed training, how much memory devices used during training, etc \cite{fed_system_design}. These metrics ensure that users' devices are not oversubscribed (draining battery) and may be used to debug device performance issues. Device analytics play a critical role in maintaining user experience quality: "Device utility to the user is mission critical, and degradations are difficult to pinpoint and easy to wrongly diagnose. Using accurate analytics to prevent federated training from negatively impacting the device’s utility to the user accounts for a substantial part of our engineering and risk mitigation costs." \cite{fed_system_design}

In our work, we leverage summary information from device analytics -- specifically how often a user performed training -- to assist disaggregating gradients, breaking privacy. Note while \cite{fed_system_design} points out that device analytics contain no personally identifiable information, these reports nevertheless provide crucial information that links gradient information collected across rounds, facilitating our attack on disaggregating gradients.

\subsection{Product Identification by Solving Linear Inverse Problem} 

Recently, independent of our work, research has shown that individual item product prices may be recovered given  customer's transaction history by optimizing a linear inverse problem \cite{linear_inverse}.  Under certain conditions (e.g: assuming that in the transaction history each item was purchased by itself at least once) their approach recovers these item prices with high precision and allows them to reveal customers' spending habits. Specifically, given a corpus of sums of item prices from customers' transaction histories, \cite{linear_inverse} utilizes a subset sum algorithm to uncover the individual prices of the transaction and to identify the products themselves.

Our work on gradient disaggregation and the work in \cite{linear_inverse} solve the same core problem: uncovering individual values given observations of their sums. While their work recovers prices of items, our work analogically reconstructs participants' model gradients. However, a key distinction in \cite{linear_inverse} is the assumption that each item must be purchased individually at least once. This makes their approach unsuitable for disaggregating aggregated model updates as, under the secure aggregation protocol, each aggregated update is composed of more than one participant's model updates. 

\subsection{Data Leakage from Model Updates}

Recent research has shown that model updates and gradients leak significant amounts of information. Information leaked by model updates ranges from specific properties to entire data samples \cite{melis2018exploiting, zhu2019deep, geiping2020inverting, shokri2017membership, qian2020learn, wei2020framework, lyu2020threats, obfuscated-gradients, mengkai_user_level, gan_inference}. Methods to recover this information from gradients are broadly categorized as inference attacks, and prior works have demonstrated the effectiveness of inference attacks on small batches of gradients, across various modalities ranging from image to text \cite{shokri2017membership}, on both shallow and deep networks \cite{geiping2020inverting}.

In the context of federated learning, these methods suffer decreased efficacy with larger aggregates ($> 100$) \cite{melis2018exploiting,zhu2019deep,geiping2020inverting}. Our work on gradient disaggregation facilitates these attacks by de-obfuscating these updates and by enabling attribution of learned properties to specific users.

\subsection{Privacy Attacks in Federated Learning}

Recent works have introduced various privacy attacks on federated learning. Broadly, these attacks are performed by a malicious central server or by participants with influence over model training \cite{lyu2020threats}. Threats from an adversarial central server typically involve extracting private information via inference attacks as described in the previous section. Attacks by adversarial participants, on the other hand, involve influencing the model training process to alter the behavior of the trained model (e.g: model poisoning, backdoors)\cite{wang2020attack, bagdasaryan2019backdoor, fung2020mitigating, byzantine_tolerant}. 

Our work on gradient disaggregation falls under the category of an attack performed by a malicious central server. Specifically, gradient disaggregation breaks the secure aggregation protocol and enables a central server to perform inference attacks on individual participants' model updates. 

\section{Gradient Disaggregation}

\subsection{Problem Statement, Threat Model and Assumptions}

Gradient disaggregation involves uncovering individual participants' model updates given observations of their sums. Concretely, on round $r$ the central server receives

$$
{G_{aggregated}}_{[r,:]} = \sum_{s \in S_{r}} M(s)
$$

where $S_{r}$ is the selected participants on round $r$, and $M(s)$ are the model updates. The goal of gradient disaggregation is, acting as an adversarial central server, to recover $M(s)$ given $G_{aggregated}$ (aggregated gradients across $n$ rounds). 

Our threat model and assumptions are as follows:
\begin{itemize}
    \item The central server is adversarial but is limited in its ability to modify the training protocol. Specifically, we assume the central server may fix its model across rounds. Such a scenario is realistic in a case where an attacker has read access to corporation servers (e.g: to collect round model update data) and limited influence over when the global model is updated (e.g: to fix the model across rounds).  An adversarial central server is a major threat model in  federated learning \cite{lyu2020threats, Li_2020,kairouz2019advances}
    \item Client selection / device participation ($S_{r}$) is somewhat random and is a subset of the total number of users. This matches the federated learning protocol which selects a random fraction of devices to participate in each round of training \cite{fed_system_design, fed_goog, Li_2020}.
    \item The central server has access to side channel information in the form of summary analytics (specifically, how often users participated across certain federated learning rounds). Device and summary analytics are a core part of federated learning systems and infrastructure \cite{fed_system_design}.
\end{itemize}

\subsection{Gradient Disaggregation by Reconstructing the User Participant Matrix}
A central server that observes aggregates of users'  updates that are constant across rounds obtains 

\begin{equation}
\begin{aligned}
G_{aggregated} = PG_{individual}
\end{aligned}
\end{equation}

where $G_{aggregated} \in \mathbb{R}^{n \times d}$ are the final aggregated dimension $d$ gradients the server collected across $n$ rounds; $P \in \{0,1\}^{n \times u}$ is the user participant matrix across the $n$ rounds with $u$ total participants specifying which users participated in which rounds; and $G_{individual} \in \mathbb{R}^{u \times d}$ contains per user individual gradients. Hence, recovering $G_{individual}$ may be viewed as a matrix factorization problem where the left term is binary.

To approach this matrix factorization problem, we start with the method introduced in \cite{matrix_fact}, which, to the best of our knowledge, is one of the only works to address matrix factorization where the left term is binary. \cite{matrix_fact} first reconstructs the binary user participant matrix $P$, then recovers $G_{individual}$ by inverting $P$ from $G_{aggregated}$. As observed by \cite{matrix_fact}, columns of $P$ lie in the image of $G_{aggregated}$. Hence, with $Nul(M)$ as the kernel of a matrix, an approach to solving this factorization problem would be to recover each column $p_k$ of $P$:
\vspace{-3mm}
\begin{equation}
\begin{aligned}
  \mbox{ Find } p_{k} \mbox{ s.t } & Nul(G_{aggregated}^T)p_{k} = 0 \\
        & p_{k} \in \{0,1\}^n
\end{aligned}
\end{equation}
\vspace{-3mm}

In the context of federated learning, this attempts to recover individually for each user which rounds they participated in. Note that such an optimization procedure can be solved using standard mixed-integer programming frameworks such as \cite{gurobi} and can additionally be parallelized across each user. 

However, this approach is not sufficient for gradient disaggregation due to three issues: 1) failure to distinguish between the numerous binary vectors in the image of $G_{aggregated}$, 2) inability to distinguish between user solutions and 3) computational difficulties due to the exponential nature of the optimization problem (recovering $p_k$ is NP hard and \cite{matrix_fact} reports only being able to solve up to $n=30$ vectors). To address these issues, we incorporate summary analytics to assist factorization.

\subsection{Leveraging Summary Analytics to Reconstruct $P$}

% {\bf MM:  Explain what index $i$ is below... just an index over multiple constraints, yes?}
We leverage summary information from device analytics as collected in \cite{fed_system_design} to assist reconstructing $P$. Specifically, summary analytics that are collected periodically by the central server log how often a specific user participated in training and can be used to narrow down $p_k$ by limiting the total number of participations across certain training rounds (see our Related Works section for details). We capture partial information on participations across rounds by introducing linear constraints: the $i$'th constraint $C_{k}^{i} \in \{0,1\}^{n}$   specifies for the $k$'th participant the training rounds for which total number of participations $c_{k}^{i}$ is known. For example, knowing that a user participated in training 3 times between rounds 1-5 and 2 times between rounds 6-10 yields $C_{1}^{1} = [1,1,1,1,1,0,0,0,0,0]$, $c_{1}^{1} = 3$, $C_{1}^{2} = [0,0,0,0,0,1,1,1,1,1]$, $c_{1}^{2} = 2$. 

We therefore add the individual constraints.  
\begin{equation}
\begin{aligned}
{C_{k}^{i}}  p_k - c_{k}^{i}= 0
\end{aligned}
\end{equation}

% TODO: Fix the C_{k} definition
\newcommand*{\horzbar}{\rule[.5ex]{2.5ex}{0.5pt}}

After collecting all $j$ constraints and counts across all users, we combine them into 
\begin{equation}
C_k = \begin{bmatrix} \horzbar & C_{k}^{1} & \horzbar \\ & \vdots    &          \\ \horzbar & C_{k}^{j} & \horzbar \end{bmatrix}, c_k = \begin{bmatrix} c_{k}^{1} \\ \vdots \\ c_{k}^j\end{bmatrix}
\end{equation}

Incorporating them into the optimization, we obtain
\begin{equation}
\begin{aligned}
\mbox{ Find } p_{k} \mbox{ s.t. } & Nul(G_{aggregated}^T)p_{k} = 0 \\
        & p_{k} \in \{0,1\}^n \\
        & C_{k} p_{k} - c_{k} = 0
\end{aligned}
\end{equation}

We note that it is possible that devices  timestamp the exact moment they perform a round of training; in this case, $P$ may be revealed directly through the  specificity of the constraints (making the disaggregation problem solvable through a simple linear regression). However, even if devices log only the total number of times they performed training (with no timestamped data) and send these analytics back to the server once every few rounds of participation, the central server may piece  together these constraints and incorporate them into the formulation above. In other words, just knowing the number of times particular users performed training and collecting this information periodically (both of which are reasonable based on \cite{fed_system_design}), the central server may obtain enough information to carry out the gradient disaggregation attack. Incorporating summary analytics into the gradient disaggregation attack is significant as it greatly reduces the problem space, allowing a solution to a previously intractable problem.

% To incorporate individual constraints, we ensure the following

% \begin{equation}
% \begin{aligned}
% {C_{k}^{i}}^T  p_k - c_{k}^{i}= 0
% \end{aligned}
% \end{equation}

% {\bf MM:  Again check equations carefully.  $p_k$ seems incorreccted stated -- it's a vector, not 0/1.  You haven't defined that $C_k$ and $c_k$ are the corresponding collection of coonstraints.  }

% {\bf MM:  I think this paragraph should be moved to the BEGINNING of section 3.3 and rewritten slightly accordingly.  This paragraph explains and motivates this approach -- it belongs at the beginning, not the end...}
%  On the topic of user device analytics, we note that it is probable that devices  timestamp the exact moment they performed a round of training; in this case, $P$ may be revealed directly through the  specificity of the constraints. However, even if devices log only the total number of times they performed training (with no timestamped data) and sends these analytics back to the server once every few rounds of participation, the central server may piece  together these constraints and incorporate them into the formulation above. In other words, just knowing the total number of times particular users performed training and collecting this piece of information periodically (both of which are reasonable based on \cite{fed_system_design}), the central server may obtain enough information to carry out this attack.
 
\subsection{Disaggregating Noisy Model Updates}
Previously, we assumed users submitted the same model update across every round. However, participants may perform updates composed of multiple steps (e.g: FedAvg) or their data may change, leading to differences in the updates they submit  across rounds. We treat these differences as a form of injected noise.

Accounting for noise, our formulation becomes
\begin{equation}
\begin{aligned}
G_{aggregated} = PG_{individual\_avg} + noise
\end{aligned}
\end{equation}

and our goal is to recover for each user the average model update they submitted across rounds $G_{individual\_avg}$. We introduce two changes to reconstruct $P$ in the presence of noise: 1) we use hard-threshold SVD with $u$ singular values to approximate the low rank product $PG_{individual\_avg}$ and 2) we relax our constraint satisfaction problem to minimize the distance of the user participant column to the image of $G_{aggregated}$:

\begin{equation}
\begin{aligned}
\mbox{ min } & ||Nul(G_{aggregated}^T)p_{k}||^2 \\
        & p_{k} \in \{0,1\}^n \\ 
        & C_{k} p_{k} - c_{k} = 0
\end{aligned}
\end{equation}

% {\bf MM:  The above works fine if the $C_k$ constraints are exact.  If they are NOT exact, should you explain you can weight to take 
% $\min |Nul(G_{aggregated})p_{k}||^2 + \lambda ||C_kp_k - c_k||^2$ for some
% parameter $\lambda$
% }

These two changes allow reconstructing $P$ even when the updates user submit across rounds are noisy. 

% Note that, as implied by the formulation, partial information is specified by sparser $C_{k}$ or with fewer constraints and enables solving settings where information is limited. 
Note that we may have incomplete information for each user;  for example, we may have constraints for a user over certain rounds but not others if the infrastructure only provides that information sporadically (or hides it).  Additionally, if round participations are inexact (e.g: off by some small error), we may relax the hard constraint $C_{k} p_{k} - c_{k} = 0$ to be a soft constraint: $min ||C_{k} p_{k} - c_{k}||^2$ and reweight the objective accordingly. Additionally, we can check whether our solution exactly recovers $P$ by probing the number of optimal solutions returned by the mixed integer programming solver; if the solver returned only one optimal solution (and proved that it is the only one), then this indicates that our reconstruction of $P$ is exact. Our full gradient disaggregation attack which works both for noisy and non-noisy updates is presented in Algorithm \ref{alg:grad_disag}.

\begin{algorithm}[h]
   \caption{Gradient Disaggregation}
   \label{alg:grad_disag}
\begin{algorithmic}
   \STATE {\bfseries Input:} Aggregated gradients $G_{aggregated}$; constraint windows $C$; constraint sums $c$, number of users $u$\\
   \STATE {\bfseries Output:} Disaggregated gradients $G_{individual\_avg}$\\
   \STATE 
   \STATE $U, \Sigma, V \longleftarrow SVD(G_{aggregated})$
   \STATE $G_{denoised} \longleftarrow U \Sigma [0:u] V$
   \FOR{$i=1$ {\bfseries to} $u$}
   \STATE $p_{i} \longleftarrow \mbox{ min } ||Nul(G_{denoised}^T)p_{i}||^2 \mbox{ s.t. } p_{i} \in \{0,1\}^n \mbox{ and } C_{i} p_{i} - c_{i} = 0$
   \ENDFOR
   \STATE
   \STATE $P \longleftarrow [p_1, ..., p_u]$ \\
   \STATE \textbf{return} $LeastSquares(P, G_{aggregated})$
   
\end{algorithmic}
\end{algorithm}

\section{Results}

\subsection{Capabilities and Limitations of Disaggregation}
 We experimentally validate the capabilities and limitations of our gradient disaggregation procedure across various parameter settings. Note that unlike prior works performing server side attacks on privacy in federated learning, our method leverages participant information and rounds of aggregated gradients. Hence in our experiments we generate this information (across various settings) to understand how our attack behaves under different conditions. We evaluate the following parameters:
 
 % TODO: 
 % - When does aggregation work vs not
 
\begin{itemize}[noitemsep]
    \item \textbf{Number of Rounds:} Number of rounds of training $n$
    \item \textbf{Number of Users:} Number of users in system $u$.
    \item \textbf{Participation Rate:} Fraction of participants chosen to participate in each round.
    \item \textbf{Constraint Granularity:} Granularity of windows across rounds with known participation sums, per user. (E.g: granularity of 10 means we know how many times each user participated across every 10 rounds).
    \item \textbf{Gradient Noise:} Noise of user model updates across rounds.
\end{itemize}
\vspace{-3mm}
We run all experiments on a 64-core cpu and use the Gurobi optimizer \cite{gurobi}.

\begin{figure}[h]
  \centering
  \subfigure[Rounds required to reconstruct $P$ with 100\% accuracy vs number of users.]{
  \includegraphics[width=.45\linewidth]{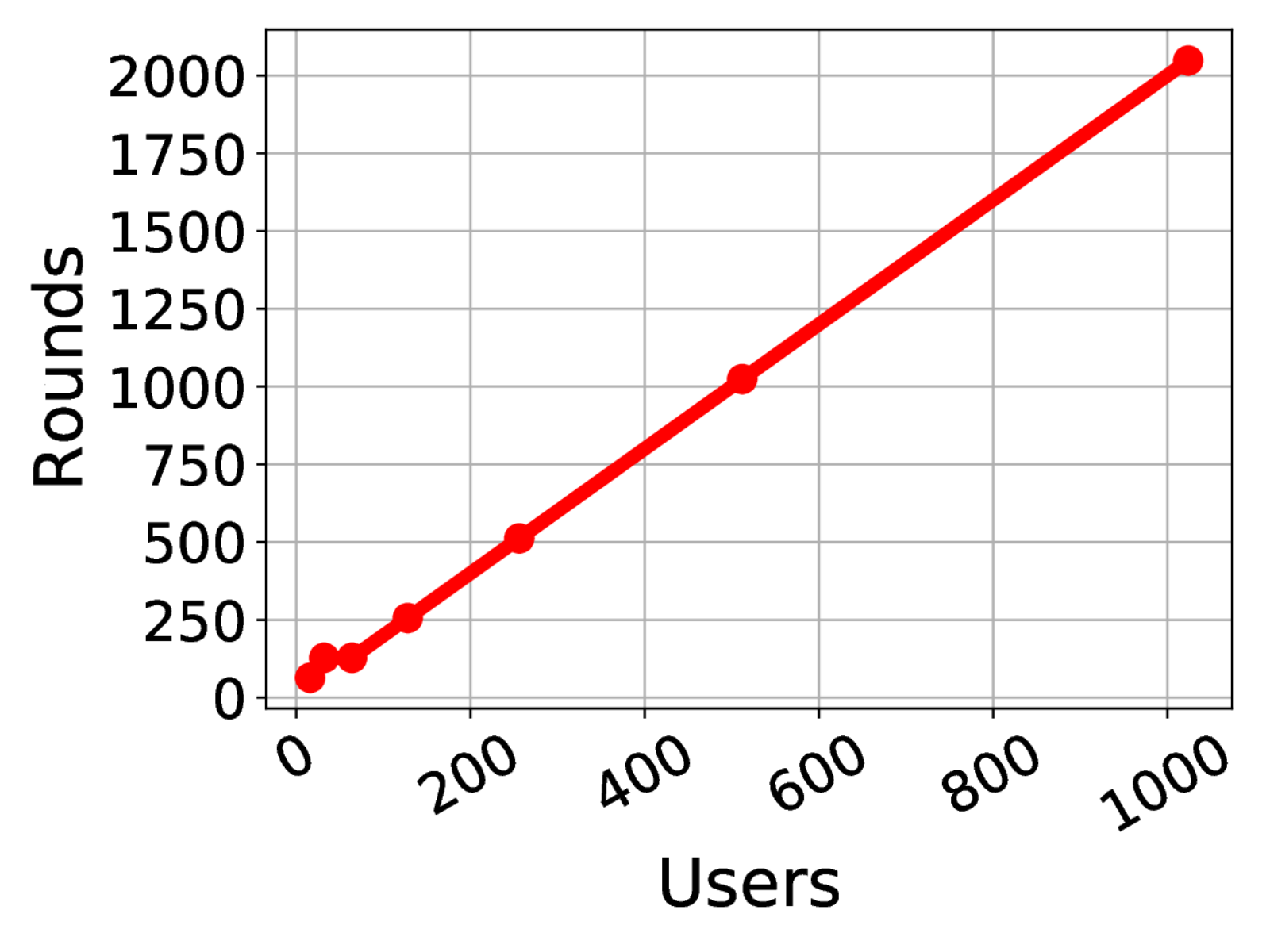}
  \label{fig:rounds_vs_users}
  }\quad
  \hfill
  \subfigure[Success rate of recovering columns of $P$ versus rounds.]{  
  \includegraphics[width=.45\linewidth]{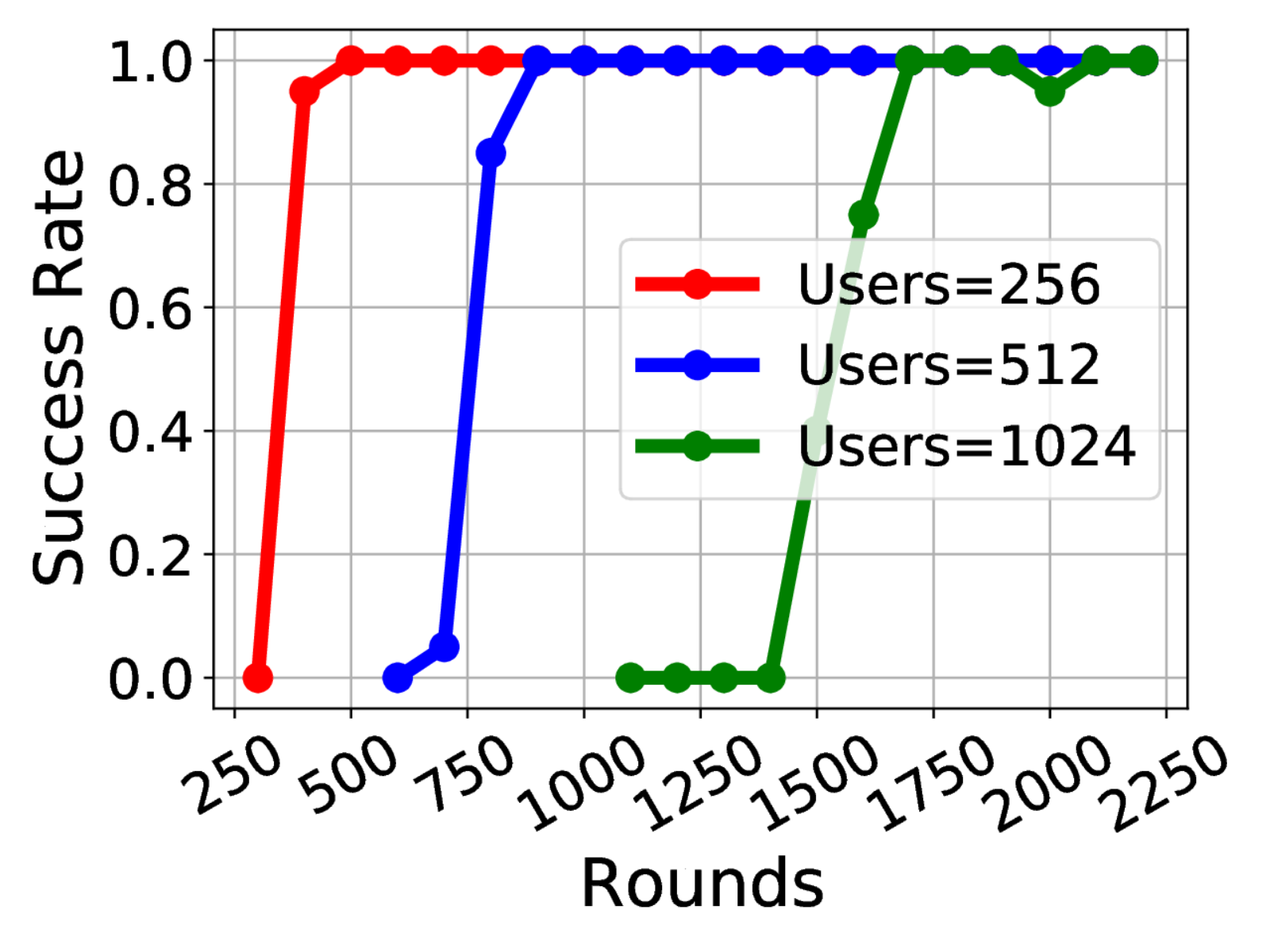}
  \label{fig:success_rate_vs_rounds}
  }
  \caption{Relationship between rounds vs number of users in gradient disaggregation. We successfully disaggregate settings with thousands of users with enough observed aggregated updates.}
\end{figure}

\begin{figure}[h]
\centering
\includegraphics[width=.45\linewidth]{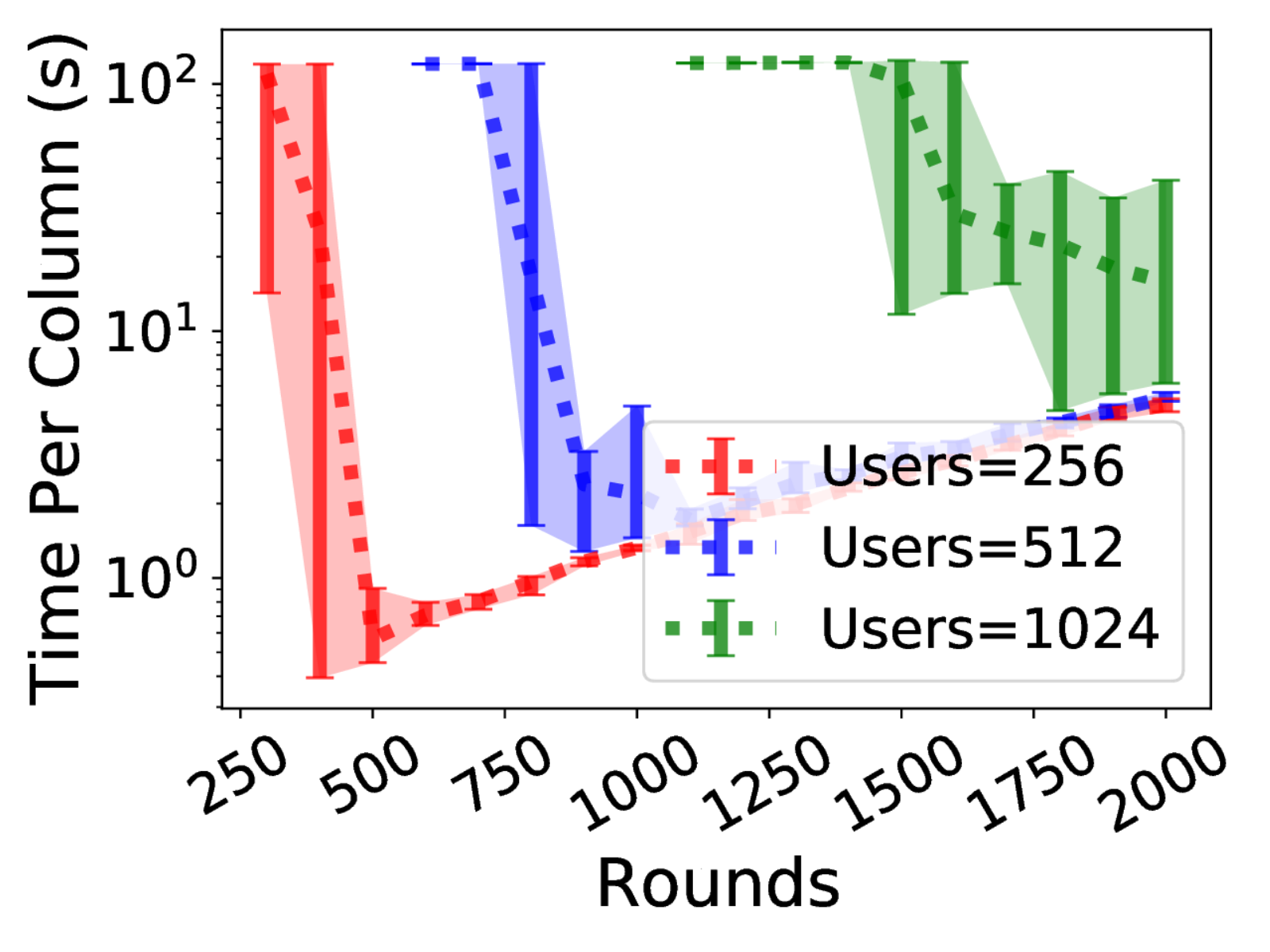}
\caption{Mean, min, and max times to reconstruct columns of $P$ vs rounds. More users require more time to recover $P$; too few rounds slows optimization due to lack of constraints; too many rounds slows optimization due to large vector sizes. We  disaggregate thousands of users' gradients in minutes on a 64-core cpu.}
\label{fig:rounds_vs_users_runtime}
\vspace{-10pt}
\end{figure}

\textbf{Number of Users}\\
We validate the maximum number of users and rounds we can disaggregate on synthetically generated matrices. $G_{individual}$ is sampled from $\mathcal{N}(0, 1)$, $P$ is sampled with sparsity = participation rate = .1, and constraint granularity=10, with no noise between submitted gradients. For users $\in$ \{16, 32, 64, 128, 256, 512, 1024\} we scan over rounds $\in$ \{16, 32, 64, 128, 256, 512, 1024, 2048\} and report the minimum number of rounds to successfully disaggregate $P$ with 100\% accuracy over 30 trials. Figure \ref{fig:rounds_vs_users} shows the number of rounds required to exactly recover $P$ across number of users; data shows that we can disaggregate matrices with thousands of user participants with enough observed gradients. Additionally, we plot success rate of reconstructing an individual column for users $\in$ \{256, 512, 1024\} which is shown in Figure \ref{fig:success_rate_vs_rounds} which furthermore reinforces that more rounds of observed gradients can increase reconstruction success rate. We also evaluate the relation that rounds vs users has on the runtime of the solver, shown in Figure \ref{fig:rounds_vs_users_runtime}, where we measure the runtime to exactly recover columns of $P$ (with a maximum time limit of 180 seconds per column). Results show that larger $P$ require more time to solve. Additionally fewer rounds leads to slower reconstruction as there are fewer constraints, while too many rounds leads to slower optimization due to large matrix sizes. Note we report time per column, as each column is solved in parallel.

\begin{figure}[h]
  \centering
  \subfigure[Rounds required to reconstruct $P$ with 100\% accuracy vs participation rate.]{
  \includegraphics[width=.45\linewidth]{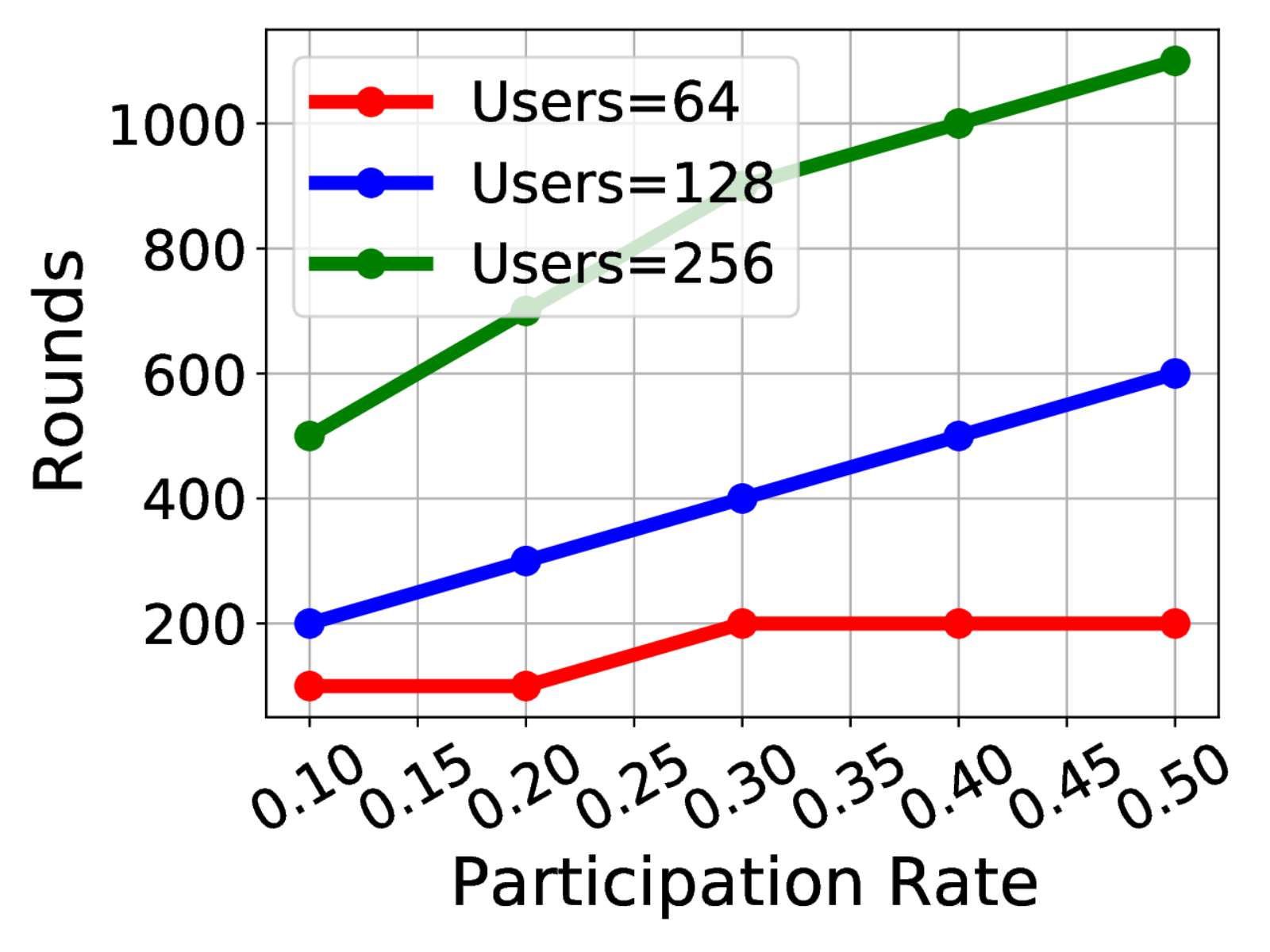}
  \label{fig:rate_vs_rounds}
  }\quad
  \hfill
  \subfigure[Mean, min, and max runtime to reconstruct columns of $P$ vs participation rate.]{  
  \includegraphics[width=.45\linewidth]{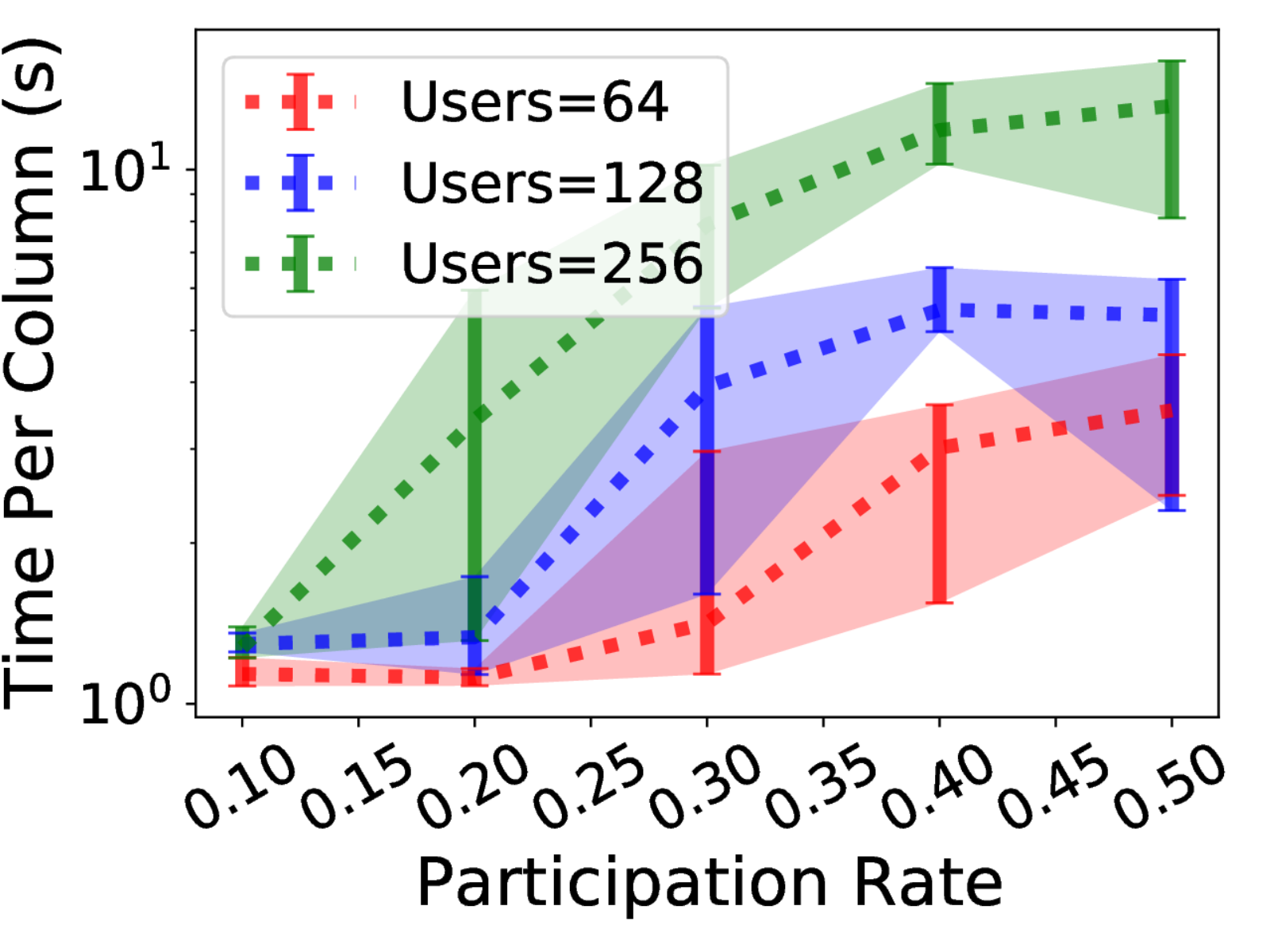}
  \label{fig:rate_vs_runtime}
  }
  \caption{Effect of participation rate in gradient disaggregation. Higher participation rate can be compensated for by observing more rounds of aggregated gradients. We recover $P$ even in the presence of many round participants.}
\end{figure}

\textbf{Participation Rate}\\
We evaluate the effect of participation rate -- the probability that a user is selected to take part in a round of training -- on gradient disaggregation. We use the same parameter settings as in the previous section  and scan participation rate $\in \{.10, .20, .30, .40, .50\}$ across various numbers of user participants, measuring number of rounds of observed aggregated gradients required to successfully reconstruct $P$ with 100\% accuracy across 30 trials. As shown in Figure \ref{fig:rate_vs_rounds}, higher participation rate requires more rounds to reconstruct $P$. Intuitively, more participants per round leads to higher obfuscation of user updates, requiring more rounds to decode. However, as indicated, by  observing more rounds of collected gradients, $P$ is eventually reconstructed exactly. We additionally evaluate participation rate's effect on runtime which is shown in \ref{fig:rate_vs_runtime}. Higher participation rate makes the reconstruction problem more difficult and hence requires longer to solve. Note that federated learning settings have between tens to hundreds of round participants \cite{Li_2020, fed_system_design} and we have chosen these points to reflect this as accurately as possible.

\begin{figure}[h]
  \centering
  \subfigure[Rounds required to reconstruct $P$ with 100\% accuracy vs constraint granularity.]{  
  \includegraphics[width=.45\linewidth]{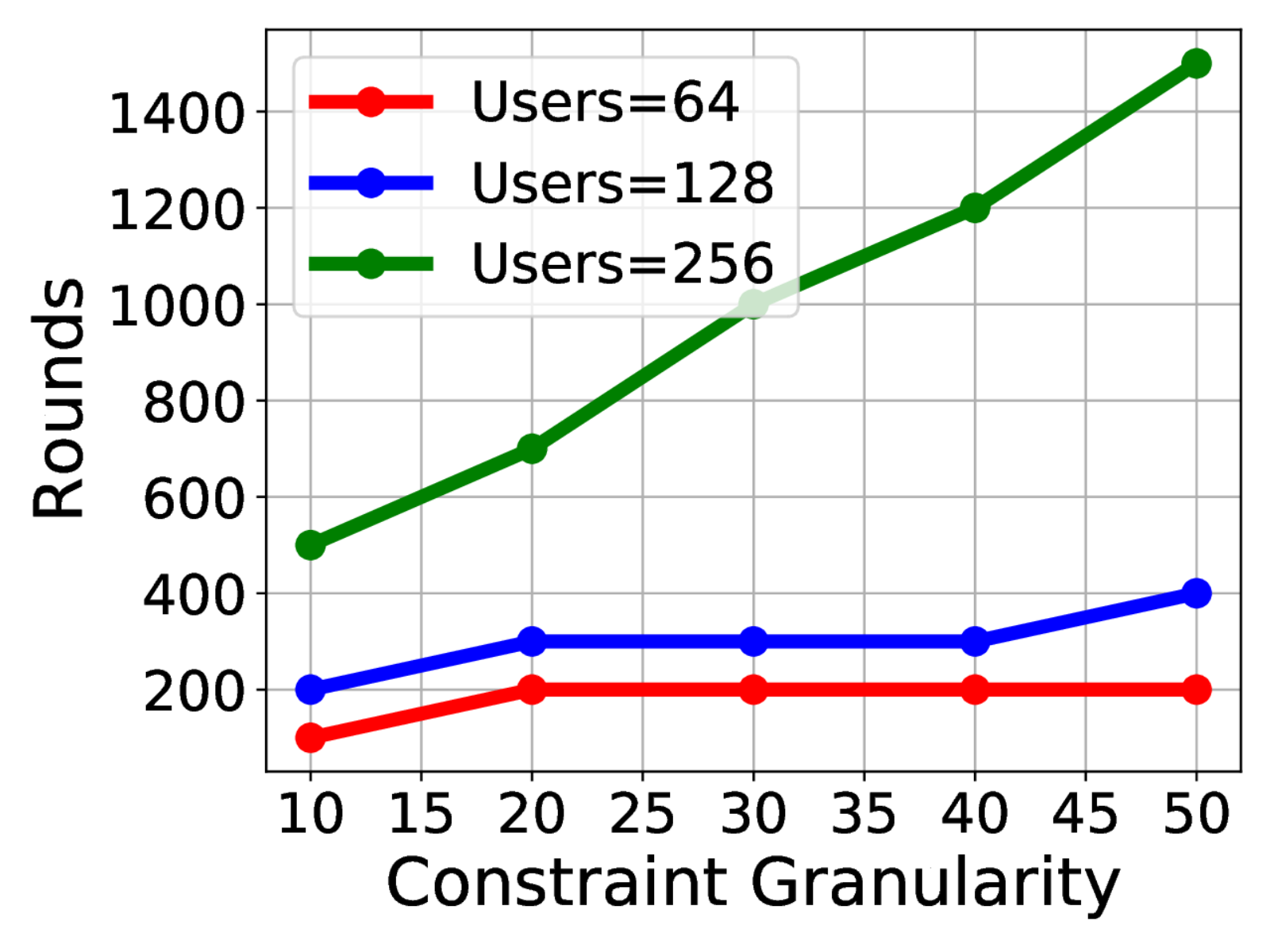}
  \label{fig:granularity_vs_rounds}
  }\quad
  \hfill
  \subfigure[Histogram of number of user participations across constraint windows.]{
  \includegraphics[width=.45\linewidth]{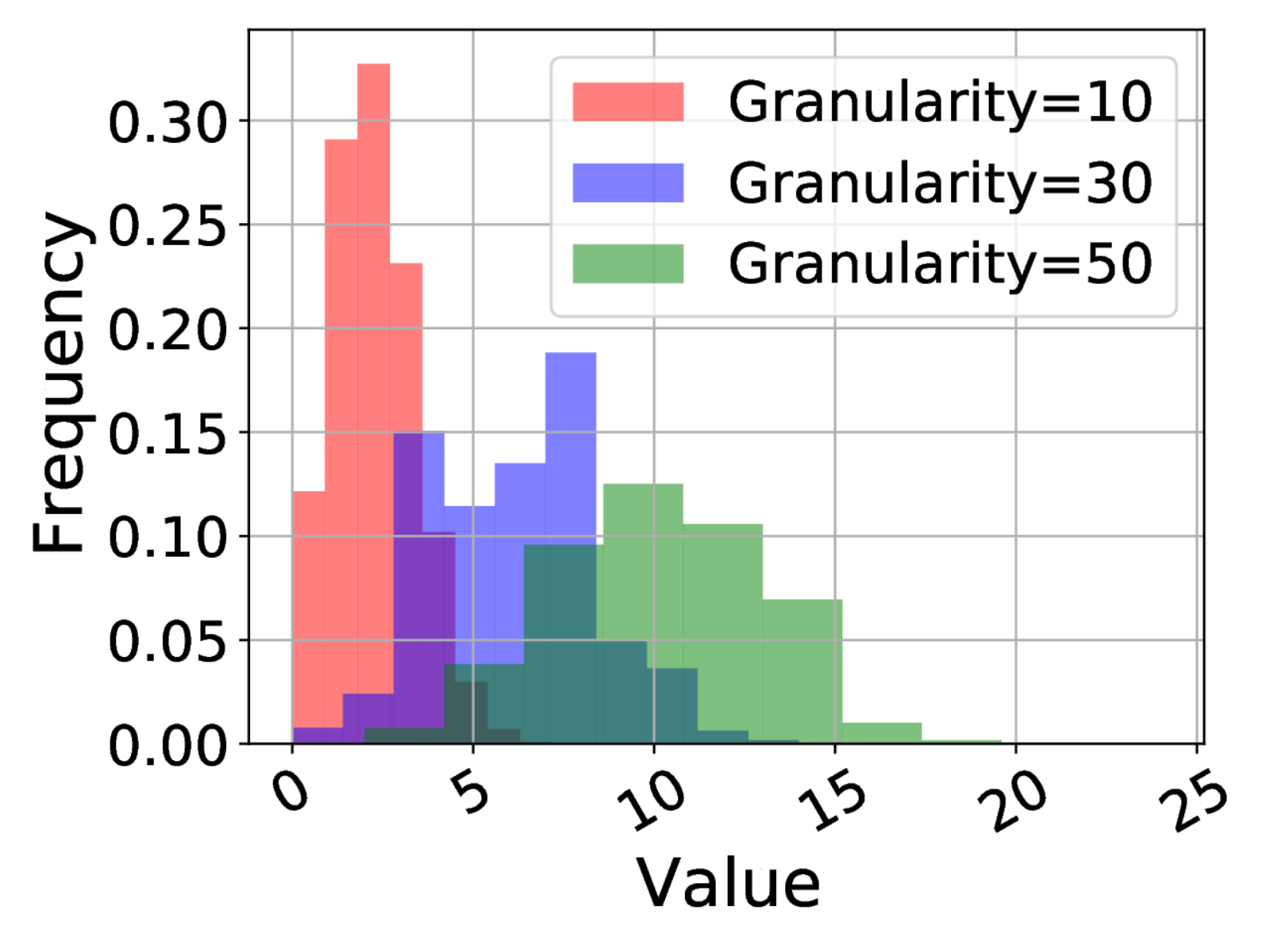}
  \label{fig:granularity_histogram}
  }  
  \caption{Effect of constraint granularity in gradient disaggregation. Coarser constraints can be compensated by observing more rounds of aggregated gradients. }
  \vspace{-5pt}
\end{figure}

\textbf{Constraint Granularity}\\
We evaluate the effect of constraint granularity on gradient disaggregation. We consider granularities $\in \{10, 20, 30, 40, 50\}$. Figure \ref{fig:granularity_vs_rounds} shows that coarser constraints make reconstruction more difficult, requiring more rounds of observed aggregated gradients. Additionally, for reference Figure \ref{fig:granularity_histogram} shows the histogram of the number of times a user participates within a granularity window at different constraint granularities. Eventually, with enough observed rounds of aggregated gradients, the participant matrix $P$ is exactly recoverable. Our results indicate that less detailed analytics may be compensated for by observing more rounds of aggregated model updates.

\textbf{Noisy Model Updates / FedAvg}\\
We address the scenario where model updates submitted by users are noisy across rounds, which may be due to the stochasticity of the optimization (e.g: the FedAvg algorithm). Initial experiments synthetically generate user ground truth gradients and inject noise into them at aggregation time. We initialize user vectors sampled from $\mathcal{N}(0,1)$ then inject noise sampled from $\mathcal{N}(0, \sigma)$ to each user's vector at aggregation time, measuring the gradient dimension required to exactly reconstruct $P$.  We perform the experiment with 100 users, a participation rate of .1 and constraint granularity of 10, with a 600 second time limit on reconstructing each column of $P$. 

\begin{figure}[h]

  \centering
  \subfigure[Number of parameters of the gradient dimension required to recover $P$ exactly.]{
      \includegraphics[width=.45\linewidth]{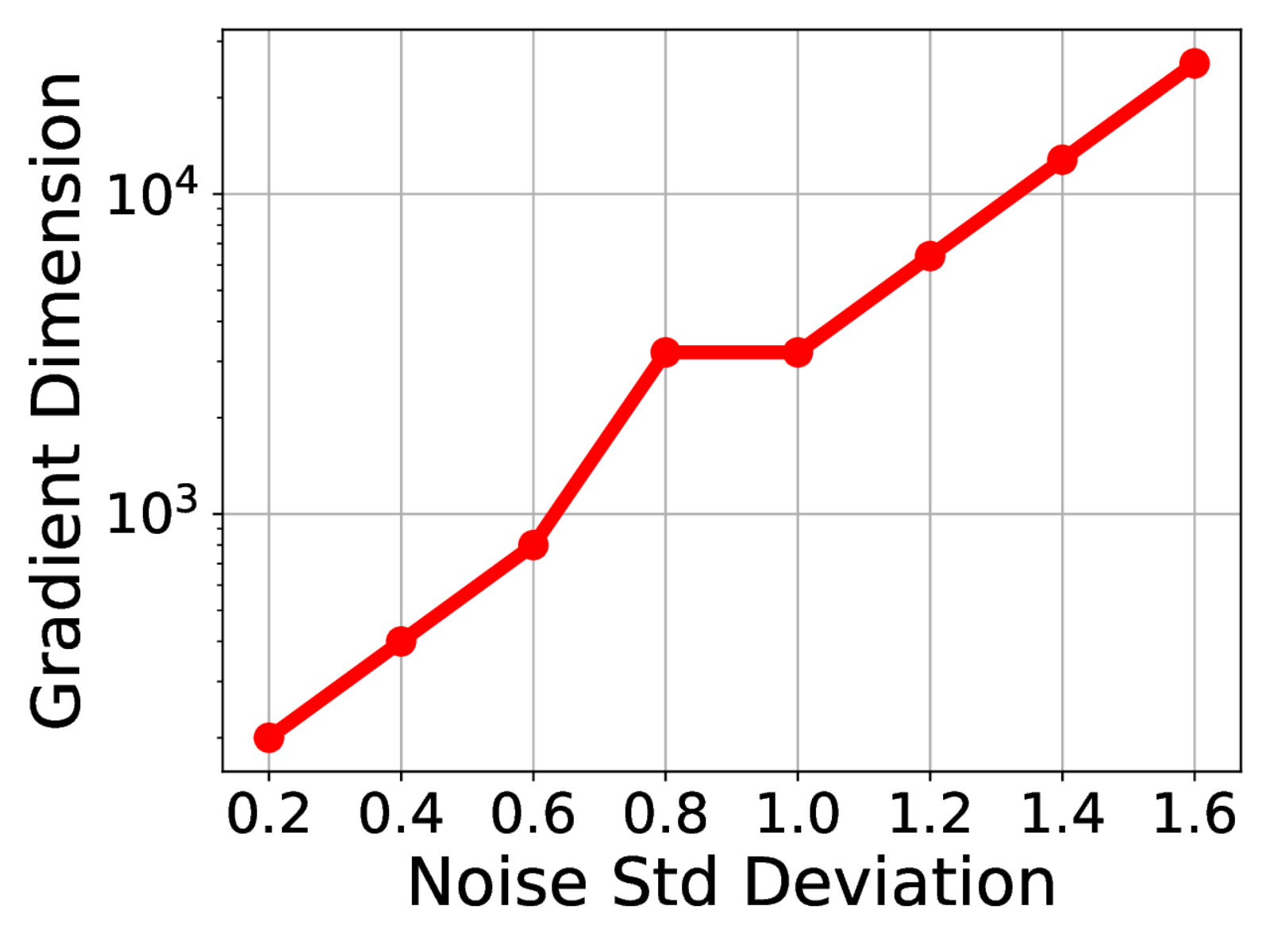}
      \label{fig:stdev_vs_dim}
  }\quad
  \hfill
  \subfigure[Relative noise of FedAvg updates on Cifar10 LeNet; batch size $b$, momentum $m$, dataset fraction $f$.]{
    \includegraphics[width=.45\linewidth]{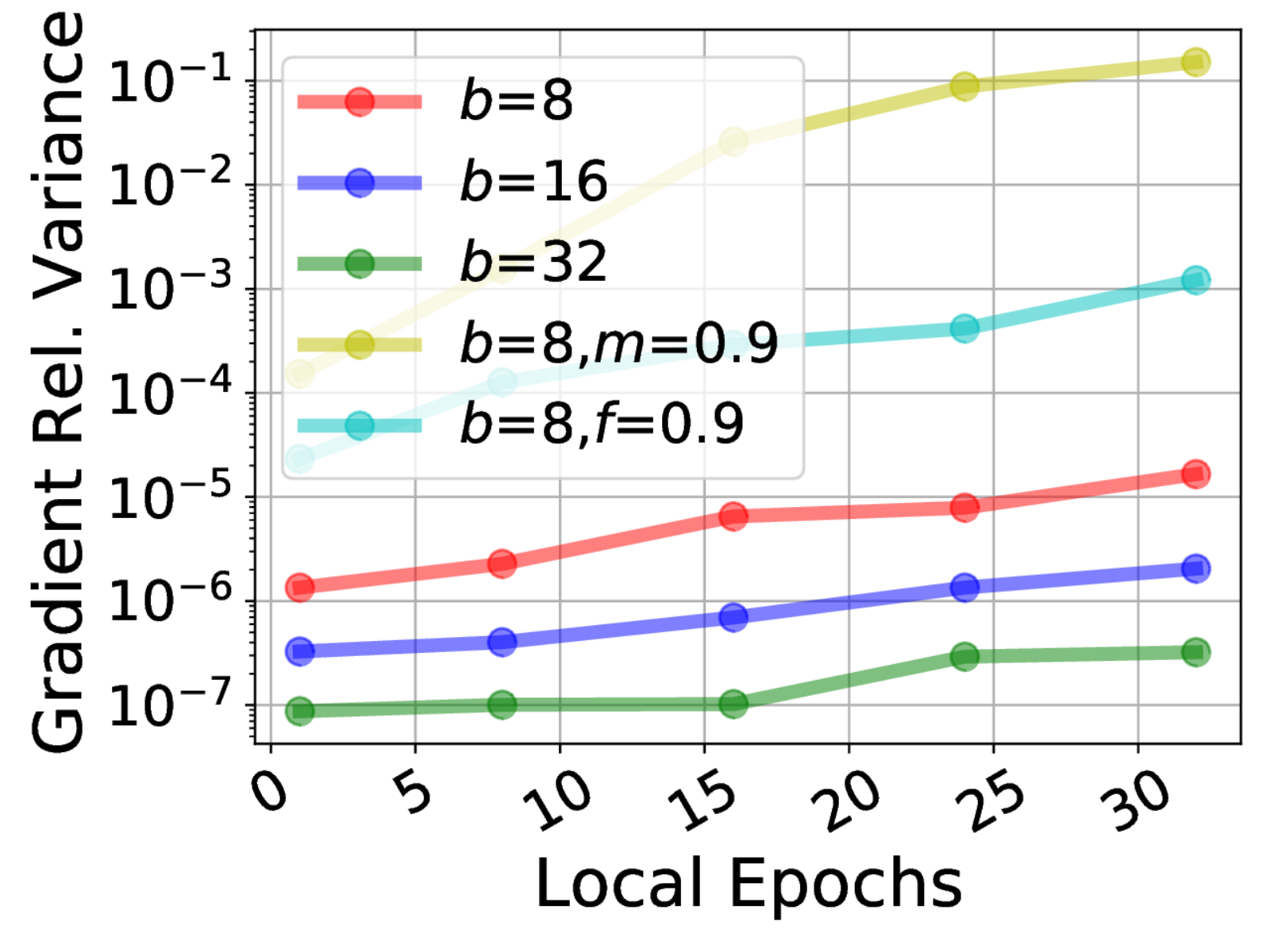}
    
     \label{fig:fedavg_variance}
  }
\caption{Effect of noise on gradient disaggregation.}
\end{figure}

 Figure \ref{fig:stdev_vs_dim} shows the minimum gradient dimension ($d$) that is required to exactly reconstruct $P$ with 100\% success rate. Note that unlike prior experiments, increased noise may be compensated for by incorporating a higher number of the parameters of the model update (rather than observing more rounds of gradients). As even the smallest neural network models contain  thousands or millions of parameters \cite{han2015deep_compression, han2015learning,howard2017mobilenets}, this indicates that the attack may handle significant levels of noise. Furthermore, note that the dimension of the model update does not significantly affect solver time as the nullspace of $G_{aggregated}$ is computed only once and reused across users.

% \usepackage{multirow}
\begin{table}[h]
\centering
\scalebox{.70}{
\begin{tabular}{|c|c|c|c|c|c|c|c|c|}
\hline
Dataset Size $D$                   & Batch Size $b$ & \multicolumn{7}{c|}{Local Epochs $e$}   \\ \hline
                                   &                & 1   & 2   & 4   & 8   & 16  & 32  & 64  \\ \hline
\multirow{3}{*}{64}                & 8              & 1.0 & 1.0 & 1.0 & 1.0 & 1.0 & 1.0 & 1.0 \\ \cline{2-9} 
                                   & 16             & 1.0 & 1.0 & 1.0 & 1.0 & 1.0 & 1.0 & 1.0 \\ \cline{2-9} 
                                   & 32             & 1.0 & 1.0 & 1.0 & 1.0 & 1.0 & 1.0 & 1.0 \\ \hline
\multirow{3}{*}{128}               & 8              & 1.0 & 1.0 & 1.0 & 1.0 & 1.0 & 1.0 & 1.0 \\ \cline{2-9} 
                                   & 16             & 1.0 & 1.0 & 1.0 & 1.0 & 1.0 & 1.0 & 1.0 \\ \cline{2-9} 
                                   & 32             & 1.0 & 1.0 & 1.0 & 1.0 & 1.0 & 1.0 & 1.0 \\ \hline
\multirow{3}{*}{64 (momentum=.9)}  & 8              & .99 & 1.0 & 1.0 & 1.0 & 1.0 & 1.0 & 1.0 \\ \cline{2-9} 
                                   & 16             & 1.0 & 1.0 & 1.0 & 1.0 & 1.0 & 1.0 & 1.0 \\ \cline{2-9} 
                                   & 32             & 1.0 & 1.0 & 1.0 & 1.0 & 1.0 & 1.0 & 1.0 \\ \hline
\multirow{3}{*}{128 (momentum=.9)} & 8              & 1.0 & 1.0 & 1.0 & 1.0 & 1.0 & 1.0 & .96 \\ \cline{2-9} 
                                   & 16             & 1.0 & 1.0 & 1.0 & 1.0 & 1.0 & 1.0 & 1.0 \\ \cline{2-9} 
                                   & 32             & 1.0 & 1.0 & 1.0 & 1.0 & 1.0 & 1.0 & 1.0 \\ \hline
\multirow{3}{*}{64 (fraction=.9)}  & 8              & .06 & .66 & .97 & .99 & .98 & .99 & .85 \\ \cline{2-9} 
                                   & 16             & 1.0 & 1.0 & 1.0 & 1.0 & 1.0 & 1.0 & .99 \\ \cline{2-9} 
                                   & 32             & 1.0 & 1.0 & 1.0 & 1.0 & 1.0 & 1.0 & 1.0 \\ \hline
\multirow{3}{*}{128 (fraction=.9)} & 8              & 1.0 & 1.0 & 1.0 & 1.0 & 1.0 & 1.0 & .90 \\ \cline{2-9} 
                                   & 16             & .59 & .96 & 1.0 & 1.0 & 1.0 & 1.0 & .99 \\ \cline{2-9} 
                                   & 32             & 1.0 & 1.0 & 1.0 & 1.0 & 1.0 & 1.0 & 1.0 \\ \hline
\end{tabular}}
    \caption{Fraction of $P$ reconstructed with FedAvg model updates (users=100, rounds=200, Cifar10 LeNet, participant rate=.1, granularity=10, time limit per column=10 min). We exactly reconstruct $P$ in the majority of FedAvg settings.}
    \label{tab:fedavg_success}
    \vspace{-5pt}
\end{table}

Additionally, we perform experiments on gradient disaggregation using model updates generated by the FedAvg algorithm \cite{mcmahan2017communicationefficient}, on Cifar10 \cite{Krizhevsky09learningmultiple} with a LeNet neural network (SGD lr=.01). FedAvg performs multiple epochs of training over the participant's dataset before sending the final model difference back to the central server. We evaluate gradient disaggregation on updates generated by FedAvg over various parameter settings: local batchsize $b$, epochs $e$, user dataset size $D$ (see \cite{mcmahan2017communicationefficient} for more details on these parameters); additionally, we simulate a shift in data distribution by randomly sampling a fraction $f$ of participants' total data set during computation of model updates; finally we test disaggregation on updates generated with and without SGD momentum $m$. Figure \ref{fig:fedavg_variance} shows that  relative variance of model updates ($D=128$) increases with epochs of training, with momentum and with a shifting data distribution. However, as Table \ref{tab:fedavg_success} shows we can reconstruct $P$ exactly in nearly all cases. The failure cases happen at lower ($\leq$ 1) or higher epochs ($\geq$ 64) of training. At lower epochs, we believe parameters of the update are smaller and less distinguished from each other, making reconstruction more difficult; at higher epochs, reconstruction is more difficult as updates are more noisy. With $2-32$ epochs, we are generally able to exactly recover $P$ across the settings.

\subsection{Gradient Inversion Attacks with Disaggregation}

We evaluate the benefits of gradient disaggregation on two methods to invert images from their gradients. Generally, gradient inversion methods optimize image data $x'$,$y'$ to match the target gradient $\nabla W$ : $\mbox{arg min}_{x',y'} || \frac{\partial l(F(x', W), y')}{\nabla W} - \nabla W||^2$ \cite{zhu2019deep}. This optimization grows exponentially more difficult with larger aggregates \cite{geiping2020inverting, zhu2019deep}; we use gradient disaggregation to reduce the aggregate and improve the quality of the inverted images. To quantitatively measure quality, we use PSNR as in \cite{geiping2020inverting}. In our results we  only show the reconstructed image with the smallest corresponding PSNR to a ground truth image for space.

\begin{figure}[H]
\centering
\subfigure[users=1 (or with disaggregation)]{
  \centering
  \includegraphics[width=.17\linewidth]{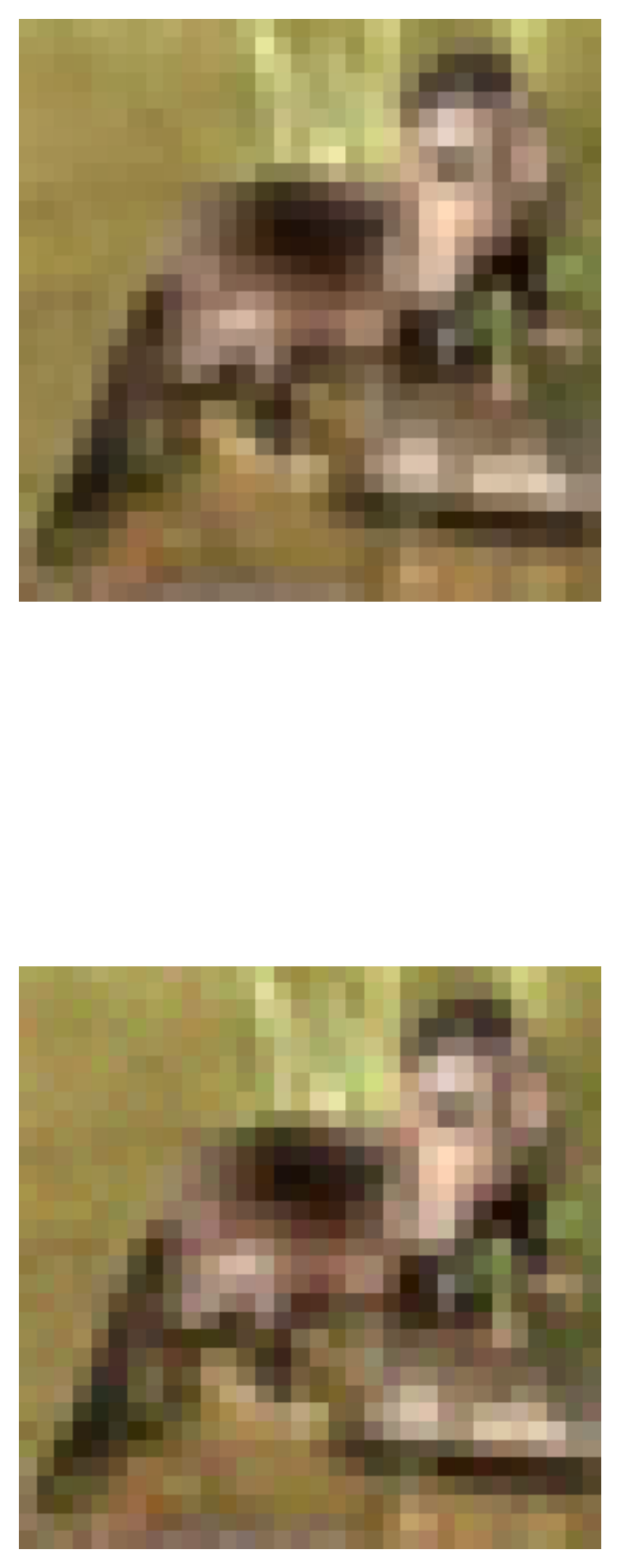}
  }\quad
  \hfill
\subfigure[users=2 (no disaggregation)]{
  \centering
  \includegraphics[width=.17\linewidth]{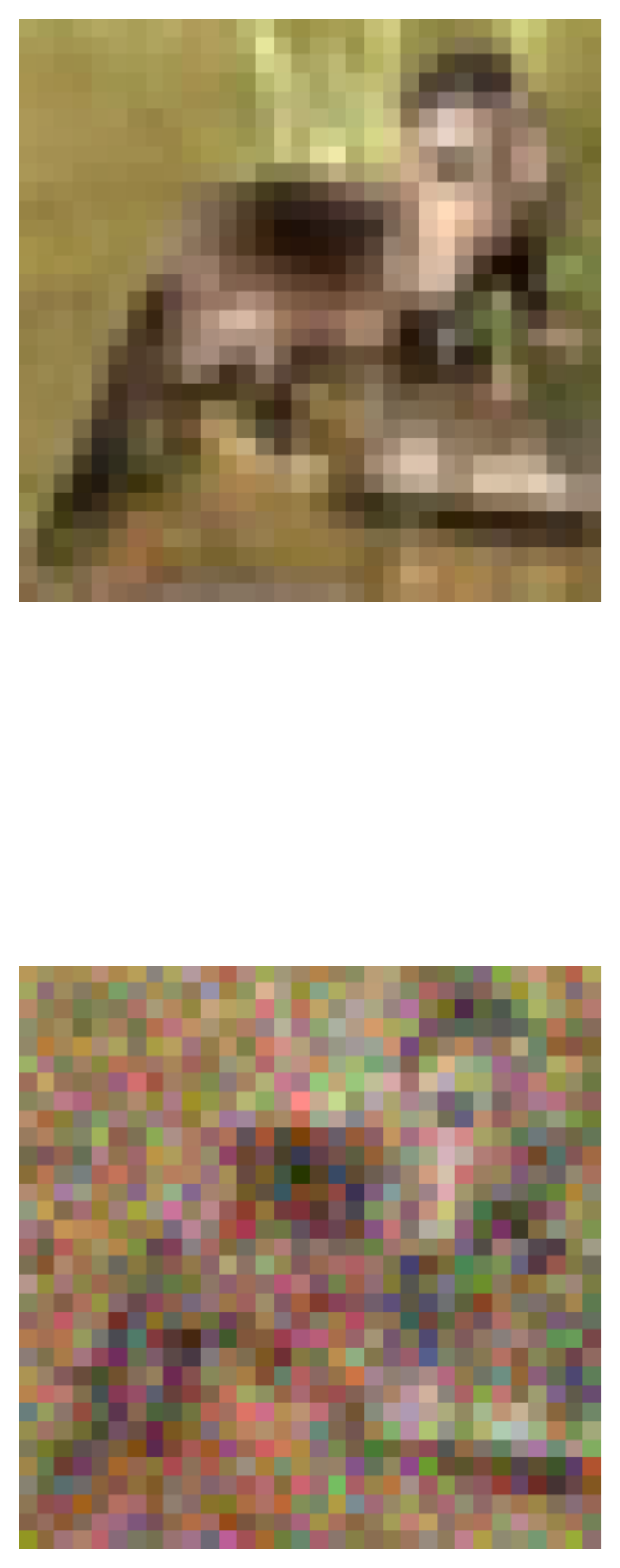}
  }\quad
  \hfill
\subfigure[users=4 (no disaggregation)]{
  \centering
  \includegraphics[width=.17\linewidth]{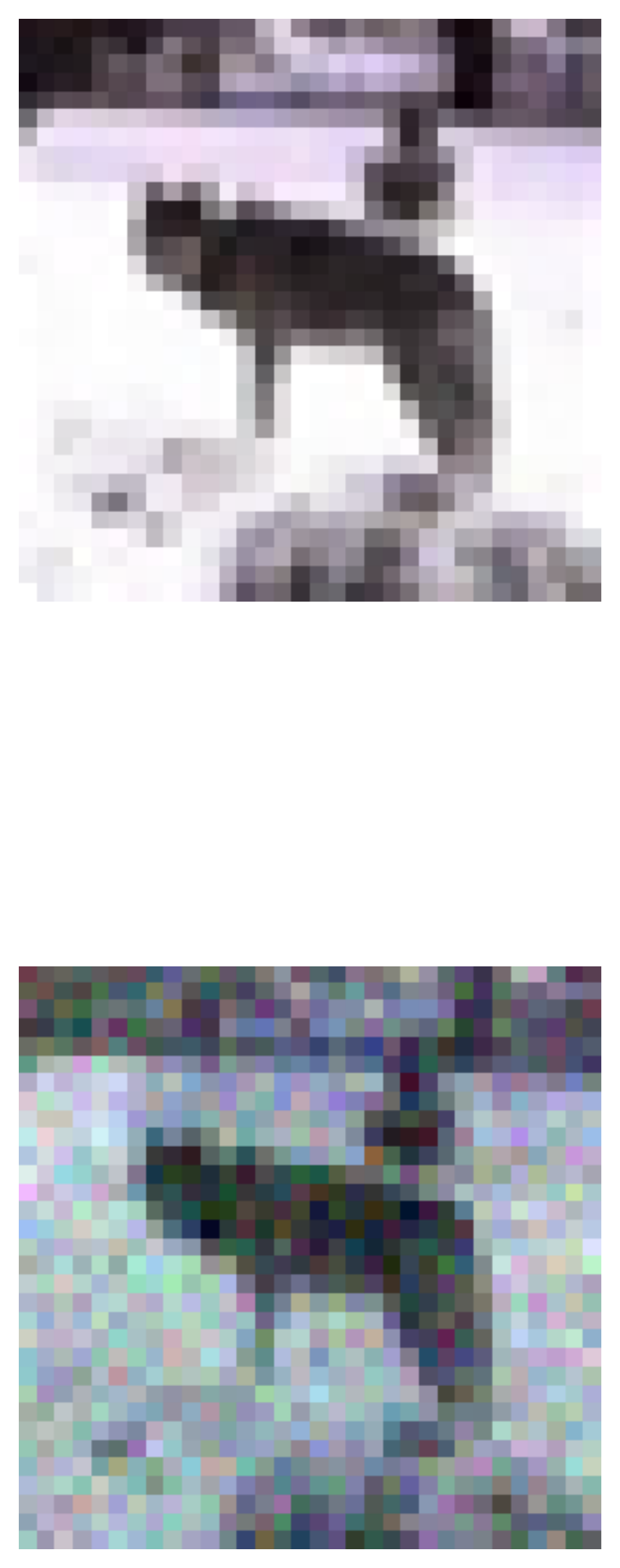}
  }\quad
  \hfill
\subfigure[users=32 (no disaggregation)]{
  \centering
  \includegraphics[width=.17\linewidth]{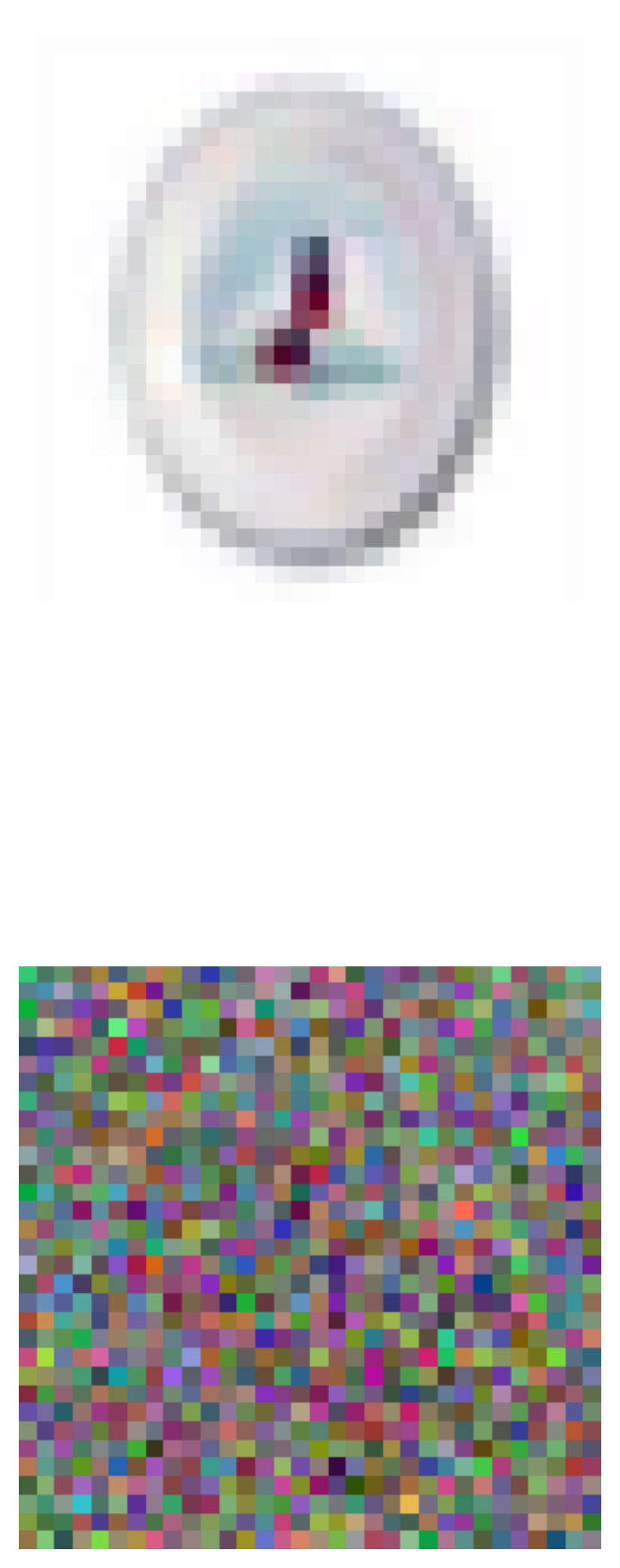}
  }\quad
  \hfill  
\caption{Recovered images from gradients across users (top image is the closest ground truth). Gradient disaggregation recovers individual users' exact gradients, hence, performing the gradient inversion attack with gradient disaggregation on multiple users yields the same quality as performing the attack on just one user. Without disaggregation, gradient inversion fails on gradients aggregated across more users.}  
\label{fig:inversion_image_cmp}
\vspace{-15pt}
\end{figure}

We perform the attack in \cite{zhu2019deep} on an MLP network on Cifar100 and show the effect of inversion with and without gradient disaggregation across multiple users with each user having 1 image in their dataset (submitting full gradients of that image).  Figure \ref{fig:inversion_image_cmp} shows the closest reconstructed image to a user's data example and Table \ref{tab:inversion_image_cmp} shows the corresponding PSNR achieved. With gradient disaggregation, we recover the target user's exact gradient and hence the reconstructed image is high quality. Without disaggregation, reconstruction quality degrades significantly.

\begin{figure}[h]
\centering
\subfigure[users=1 (no disaggregation)]{
  \centering
  \includegraphics[width=.25\linewidth]{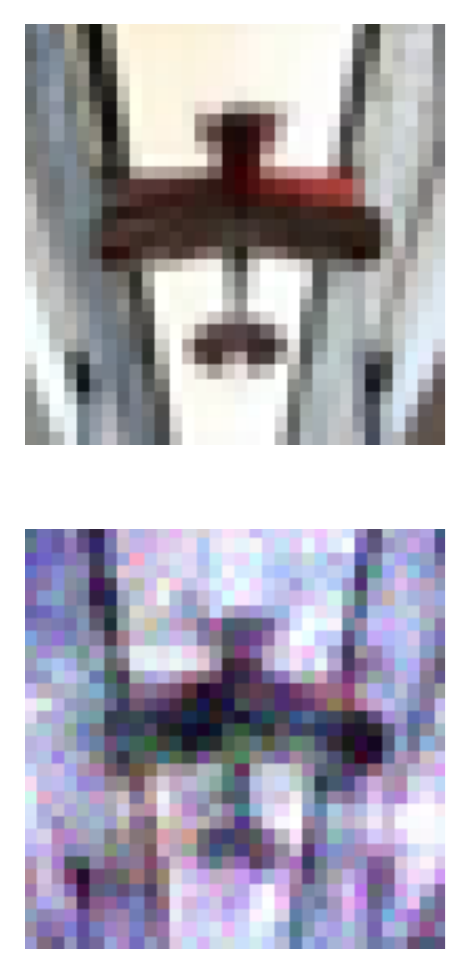}
  }\quad
  \hfill
\subfigure[users=10 (no disaggregation)]{
  \centering
  \includegraphics[width=.25\linewidth]{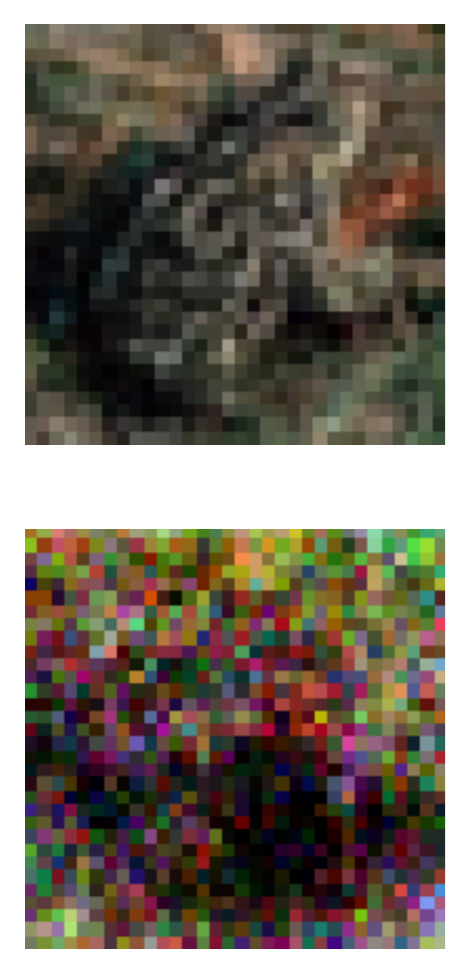}
  }\quad
  \hfill  
\subfigure[users=100 (disaggregated)]{
  \centering
  \includegraphics[width=.25\linewidth]{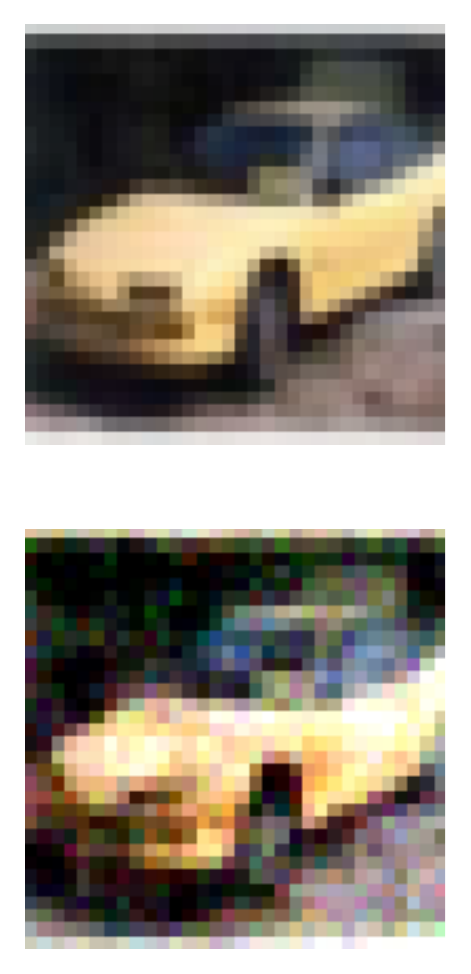}
  }\quad
  \hfill

\caption{Recovered images from FedAvg updates across users (top image is closest ground truth). Gradient disaggregation enables high quality inversion on noisy FedAvg updates aggregated across many users; unlike disaggregation on exact gradients, disaggregation on noisy updates recovers the average update submitted across rounds, and we are able to reconstruct high quality images on noisy updates aggregated across many users. Without disaggregation, inversion on updates aggregated over multiple users (users=10) significantly degrades quality.}

\label{fig:inversion_fedavg}
\vspace{-10pt}
\end{figure}

We furthermore perform the attack in \cite{geiping2020inverting} to invert noisy FedAvg updates. Figure \ref{fig:inversion_fedavg} and Table \ref{tab:inversion_fedavg} show the results of inverting fedavg updates with local epochs = 4, batch size = 16, user data set size = 64, with and without gradient disaggregation (100 users, 2 layer MLP). With gradient disaggregation we achieve similar quality as inverting a single model update, whereas inverting an update aggregated over multiple users (users=10) significantly degrades reconstruction quality. %Note that with gradient disaggregation the reconstructed image is attributed to that particular user; without it, a reconstructed image may have been from any of the users.

\begin{table}[H]
\centering
\scalebox{.8}{
\begin{tabular}{|c|c|c|c|c|}
\hline
     & users=1  & users=2  & users=4  & users=32 \\ \hline
PSNR & 36.5 & 18.8 & 13.9 & 6.1  \\ \hline
\end{tabular}
}

\caption{Corresponding PSNR scores against ground truth for Figure \ref{fig:inversion_image_cmp}}
\label{tab:inversion_image_cmp}
\vspace{-25pt}

\end{table}

\begin{table}[h]
\centering
\scalebox{.8}{
\begin{tabular}{|c|c|c|c|}
\hline
     & users=1 & users=10 & \shortstack{users=100 \\ (disaggregated)}  \\ \hline
PSNR & 16.0    & 13.3 & 18.6                      \\ \hline
\end{tabular}
}
\caption{Corresponding PSNR scores against ground truth for Figure \ref{fig:inversion_fedavg}.}
\label{tab:inversion_fedavg}
\vspace{-10pt}
\end{table}

\subsection{Property Inference Attacks with Disaggregation}
We demonstrate gradient disaggregation on property inference attacks as in \cite{melis2018exploiting}. We train a gender model on the LFW dataset \cite{LFWTech} and a model to predict whether participants' FedAvg updates (local epochs=4, batchsize=8, data size per user=32) on the gender model contain people of a specific race (hence the attacker's goal may be to learn a participants' images' race from the application). As in \cite{melis2018exploiting} only the target's dataset contains a significant proportion ($p$=.5) of images with the specific race and the goal is to determine whether the target's update is present in the aggregated updates over various numbers of users.

\begin{figure}[h]
\centering
\includegraphics[width=.50\linewidth]{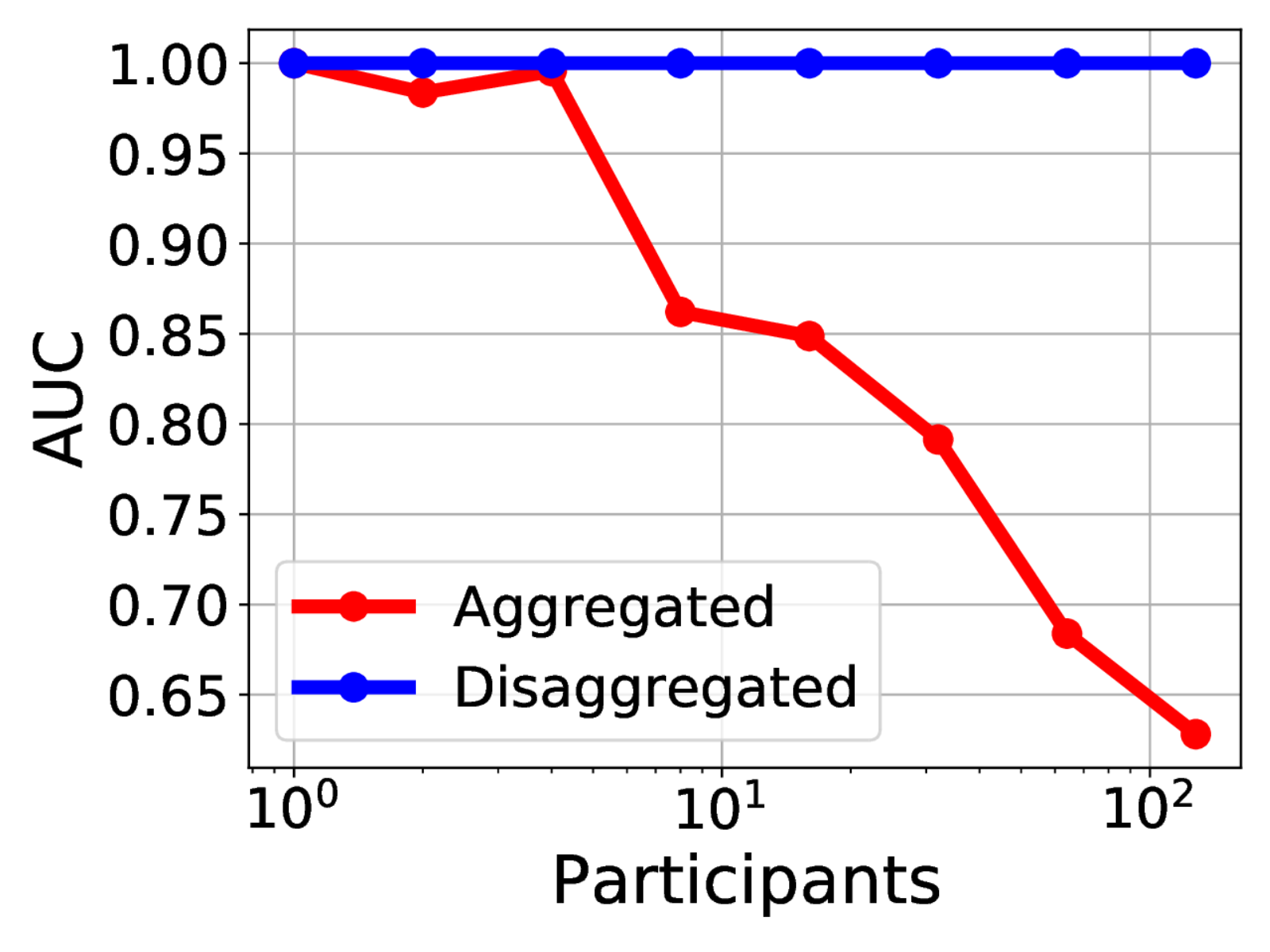}
\caption{Property inference  with and without gradient disaggregation (main task: gender classification, auxiliary task: identifying images of specific race) on FedAvg updates. Gradient disaggregation enables property inference on individual model updates and maintains high AUC score across increased number of users.}
\label{fig:property_inf}
\vspace{-10pt}

\end{figure}

Figure \ref{fig:property_inf} shows the AUC score of the attack across various numbers of users with and without gradient disaggregation. AUC score quickly degrades with more users; however, with gradient disaggregation high AUC score is maintained across increased numbers of participants as each user's model update is disaggregated exactly, allowing the property inference attack to be  performed on each user separately. We note that the requirement in \cite{melis2018exploiting} that only the target has the particular data distribution is a limiting assumption, as many participants' data may exhibit the property of interest. With gradient disaggregation, learned properties are attributed to individual participants, enabling the central server to build  profiles of users, violating anonymity.

\section{Discussion}
 % TODO: 
 % - When does aggregation work vs not
 % - Look more into analytics in fed systems

We introduce gradient disaggregation, a method to disaggregate model updates from sums of model updates given repeated observations and access to summary information from device analytics. Our attack is capable of disaggregating model updates over thousands of users and we apply it to augment existing attacks such as gradient inversion and property inference. Our attack undermines the secure aggregation protocol.

Our findings show that summary metrics such as participation frequency may, when combined with gradient information, be used as an attack vector to undermine individual users' data privacy in federated learning systems. Ways to mitigate this attack include: injecting noise into model updates to reduce efficacy of disaggregation, using differential privacy on the collected device metrics to make reconstruction more difficult, and reducing or eliminating the collection of device analytics. These mitigation strategies may hinder the management of federated learning systems, and employing these techniques to increase privacy must be balanced with the costs to utility. We hope that bringing awareness to the privacy risks of side channel information in federated learning infrastructure will assist in designing secure federated learning systems.

%This suggests that infrastructure and analytics to support federated learning systems must be designed carefully, with privacy in mind. More broadly, it indicates a need for a more wholistic view on privacy, where both the algorithm and surrounding system/infrastructure are jointly designed to ensure privacy. We believe further research in this area will lead to more secure learning systems.

\section{Acknowledgements}
We are grateful for the helpful discussions with members of Harvard Edge Computing group and VLSI group. This work was supported by the Application Driving Architectures (ADA) Research Center, a JUMP Center cosponsored by SRC and DARPA. Michael Mitzenmacher was supported in part by NSF grants CCF-1563710, CCF-1535795, and DMS-2023528, and by a gift to the Center for Research on Computation and Society at Harvard University. Maximilian Lam was supported by the Ashford Fellowship. 

\bibliography{refs}
\bibliographystyle{icml2021}

\newpage

\beginsupplement
\onecolumn

\section{Experiments on "Honest but Curious" Disaggregation Attack (neural network is updated, not fixed)}

Prior experiments assumed that the central server fixed the neural network model. In the following experiments, we eliminate this assumption and update the neural network model with model updates computed from participants (via FedAvg). In this scenario, the attack becomes honest but curious as the attacker no longer has to modify the learning protocol (specifically, via fixing the model), but instead only needs to observe the gradient information and summary analytics collected by the server.

\begin{table}[h]
\centering
\scalebox{.70}{
\begin{tabular}{|c|c|c|c|c|c|c|c|}
\hline
Dataset Size $D$     & Batch Size $b$ & \multicolumn{6}{c|}{Local Epochs $e$} \\ \hline
                     &                & 1    & 2    & 4    & 8    & 16  & 32  \\ \hline
\multirow{3}{*}{64}  & 8              & 1.0  & 1.0  & .30  & 0.0  & 0.0 & 0.0 \\ \cline{2-8} 
                     & 16             & 1.0  & 1.0  & 1.0  & .35  & 0.0 & 0.0 \\ \cline{2-8} 
                     & 32             & 1.0  & 1.0  & 1.0  & 1.0  & .35 & 0.0 \\ \hline
\multirow{3}{*}{128} & 8              & 1.0  & .52  & 0.0  & 0.0  & 0.0 & 0.0 \\ \cline{2-8} 
                     & 16             & 1.0  & 1.0  & .46  & 0.0  & 0.0 & 0.0 \\ \cline{2-8} 
                     & 32             & 1.0  & 1.0  & 1.0  & .49  & 0.0 & 0.0 \\ \hline
\end{tabular}}
\caption{"Honest but curious" gradient disaggregation on FedAvg updates: neural network model is updated with participants' updates via FedAvg (SGD lr=1e-3) every round. Even with extra added noise from the changing model, our gradient disaggregation attack may reconstruct $P$ exactly in a good portion of settings.}
\label{tab:update_model_fedavg}
\end{table}

We run our experiments using FedAvg model updates, on Cifar10, with 100 users, 200 rounds, participation rate of .1, constraint granularity of 10, SGD lr of 1e-3, on a LeNet CNN model. Table \ref{tab:update_model_fedavg} shows the fraction of $P$ reconstructed across various FedAvg settings. Results show that with lower lr (1e-3), gradient disaggregation can exactly reconstruct $P$ when FedAvg updates are small (1-4 epochs). With more epochs of FedAvg (and larger dataset size or smaller batch size), increased noise prevents reconstruction of $P$. We additionally tried the experiment with higher lr (1e-2), and disaggregation typically failed due to high noise, even with lower epochs of FedAvg. Our results indicate that under the right set of circumstances, the gradient disaggregation attack can be used in an honest but curious scenario; however, more robust results are achieved if the attacker can fix the neural network model.

\section{Experiments on Partial Constraints}

We perform experiments showing our gradient disaggregation attack with partial sets of constraints. Specifically, constraint granularity determines the rounds across which number of participations is known (e.g: if granularity is 10, we know how many times each user participated between rounds 0-9, 10-19, 20-29, etc) and we drop a specific fraction of these constraints (e.g: in prior setting, only knowing participation counts between rounds 0-9, 40-49, etc). Testing this scenario shows the degree to which our method works when only partial summary analytics is given. We enforced a time limit of 10 minutes for solving each column

\begin{table}[h]
\centering
\scalebox{.70}{
\begin{tabular}{|c|c|c|c|c|c|c|c|c|c|c|c|}
\hline
Users                 & Rounds & \multicolumn{10}{c|}{Constraint Fraction}               \\ \hline
                      &        & 1 & .9  & .8  & .7  & .6  & .5  & .4  & .3  & .2  & .1  \\ \hline
\multirow{5}{*}{128}  & 256    & 1 & 1   & 1   & 1   & 1   & .97 & .71 & .26 & .08 & .03 \\ \cline{2-12} 
                      & 512    & 1 & 1   & 1   & 1   & 1   & 1   & 1   & 1   & .99 & .38 \\ \cline{2-12} 
                      & 1024   & 1 & 1   & 1   & 1   & 1   & 1   & 1   & 1   & 1   & .99 \\ \cline{2-12} 
                      & 2048   & 1 & 1   & 1   & 1   & 1   & 1   & 1   & 1   & 1   & 1   \\ \cline{2-12} 
                      & 4096   & 1 & 1   & 1   & 1   & 1   & 1   & 1   & 1   & 1   & 1   \\ \hline
\multirow{5}{*}{256}  & 256    & 0 & 0   & 0   & 0   & 0   & 0   & 0   & 0   & 0   & 0   \\ \cline{2-12} 
                      & 512    & 1 & 1   & 1   & .98 & .94 & .63 & .09 & .02 & 0   & 0   \\ \cline{2-12} 
                      & 1024   & 1 & 1   & 1   & 1   & 1   & 1   & 1   & .99 & .70 & .10 \\ \cline{2-12} 
                      & 2048   & 1 & 1   & 1   & 1   & 1   & 1   & 1   & 1   & 1   & .96 \\ \cline{2-12} 
                      & 4096   & 1 & 1   & 1   & 1   & 1   & 1   & 1   & 1   & 1   & 1   \\ \hline
\multirow{5}{*}{512}  & 256    & 0 & 0   & 0   & 0   & 0   & 0   & 0   & 0   & 0   & 0   \\ \cline{2-12} 
                      & 512    & 0 & 0   & 0   & 0   & 0   & 0   & 0   & 0   & 0   & 0   \\ \cline{2-12} 
                      & 1024   & 1 & .99 & .99 & .93 & .53 & .09 & 0   & 0   & 0   & 0   \\ \cline{2-12} 
                      & 2048   & 1 & 1   & 1   & 1   & 1   & 1   & 1   & .99 & .55 & .01 \\ \cline{2-12} 
                      & 4096   & 1 & 1   & 1   & 1   & 1   & 1   & 1   & 1   & 1   & .85 \\ \hline
\multirow{5}{*}{1024} & 256    & 0 & 0   & 0   & 0   & 0   & 0   & 0   & 0   & 0   & 0   \\ \cline{2-12} 
                      & 512    & 0 & 0   & 0   & 0   & 0   & 0   & 0   & 0   & 0   & 0   \\ \cline{2-12} 
                      & 1024   & 0 & 0   & 0   & 0   & 0   & 0   & 0   & 0   & 0   & 0   \\ \cline{2-12} 
                      & 2048   & 1 & 1   & .99 & .95 & .41 & .01 & 0   & 0   & 0   & 0   \\ \cline{2-12} 
                      & 4096   & 1 & 1   & 1   & 1   & 1   & 1   & .99 & .81 & .1  & 0   \\ \hline
\end{tabular}}

\caption{Fraction of $P$ recovered with gradient disaggregation on partial constraint information. With more rounds, we can exactly recover $P$ even when a significant fraction of constraints are missing.}
\label{tab:constraint_fraction}
\end{table}

Table \ref{tab:constraint_fraction} shows the fraction of $P$ reconstructed across various proportions of dropped constraints. Results indicate that even when significant proportions of constraints are dropped, $P$ may be exactly recovered with more rounds of collected aggregated updates. Note that, when rounds $<$ users reconstruction fails due to the rank being less than the number of users.

\section{Experiments on Inexact "Noisy" Constraints} 

We perform experiments showing our gradient disaggregation attack when the constraints are inexact. For example, if analytics specified that a user participated 5 times between rounds 0-10, when the user actually participated 4 times. To handle noisy constraints, we relax our formulation and convert participation constraints into soft constraints:

\begin{equation}
\begin{aligned}
\mbox{ min } ||Nul(G_{aggregated}^T)p_{k}||^2  + & \lambda(C_{k} p_{k} - c_{k}) \\
        & p_{k} \in \{0,1\}^n \\ 
\end{aligned}
\end{equation}

Where $\lambda$ is a reweighting factor for the participation constraints.

Table \ref{tab:inexact_constraints} shows the results of gradient disaggregation when constraint noise $\mathcal{N}(0, \mu)$ is added to each constraint. As indicated, even in the presence of noise, gradient disaggregation may exactly recover $P$ with enough rounds.

% \usepackage{multirow}
\begin{table}[h]
\centering
\scalebox{.7}{
\begin{tabular}{|c|c|c|c|c|}
\hline
Number of Users      & Rounds & \multicolumn{3}{c|}{Constraint Noise} \\ \hline
\multirow{6}{*}{32}  &        & .3         & .5         & 1           \\ \cline{2-5} 
                     & 128    & 1          & 1          & .63         \\ \cline{2-5} 
                     & 256    & 1          & 1          & .97         \\ \cline{2-5} 
                     & 512    & 1          & 1          & 1           \\ \cline{2-5} 
                     & 1024   & 1          & 1          & 1           \\ \cline{2-5} 
                     & 2048   & 1          & 1          & 1           \\ \hline
\multirow{5}{*}{64}  & 128    & 1          & 1          & .34         \\ \cline{2-5} 
                     & 256    & 1          & 1          & .97         \\ \cline{2-5} 
                     & 512    & 1          & 1          & 1           \\ \cline{2-5} 
                     & 1024   & 1          & 1          & 1           \\ \cline{2-5} 
                     & 2048   & 1          & 1          & 1           \\ \hline
\multirow{5}{*}{128} & 128    & 0          & 0          & 0           \\ \cline{2-5} 
                     & 256    & 1          & 1          & .7          \\ \cline{2-5} 
                     & 512    & 1          & 1          & .99         \\ \cline{2-5} 
                     & 1024   & 1          & 1          & 1           \\ \cline{2-5} 
                     & 2048   & 1          & 1          & 1           \\ \hline
\multirow{5}{*}{256} & 128    & 0          & 0          & 0           \\ \cline{2-5} 
                     & 256    & 0          & 0          & 0           \\ \cline{2-5} 
                     & 512    & 1          & 1          & .48         \\ \cline{2-5} 
                     & 1024   & 1          & 1          & 1           \\ \cline{2-5} 
                     & 2048   & 1          & 1          & 1           \\ \hline
\end{tabular}}
\caption{Fraction of $P$ recovered with gradient disaggregation when constraints are inexact/noisy. Participation rate=.1, $\lambda$=.1, constraint granularity = 10. Reconstruction is more successful with more rounds.}
\label{tab:inexact_constraints}
\end{table}

\section{Extended Experiments on Participation Rate}

We provide extended data on gradient disaggregation against various parameter values of participation rate. Participation rate is the proportion of users that participate in sending model updates per round and impacts how many updates are summed to yield the aggregated update that is observed by the central server. Unless stated, we enforced a time limit of 10 minutes for solving each column; we use a constraint granularity of 10. We show our extended results in Table \ref{tab:extended_participation_rate}.

\begin{table}[H]
\centering
\scalebox{.70}{
\begin{tabular}{|c|c|c|c|c|}
\hline
Users                 & Rounds & \multicolumn{3}{c|}{Participation Rate} \\ \hline
                      &        & .1          & .2          & .4          \\ \hline
\multirow{5}{*}{128}  & 256    & 1           & 1           & .05         \\ \cline{2-5} 
                      & 512    & 1           & 1           & 1           \\ \cline{2-5} 
                      & 1024   & 1           & 1           & 1           \\ \cline{2-5} 
                      & 2048   & 1           & 1           & 1           \\ \cline{2-5} 
                      & 4096   & 1           & 1           & 1           \\ \hline
\multirow{5}{*}{256}  & 256    & 0           & 0           & 0           \\ \cline{2-5} 
                      & 512    & 1           & .73         & 0           \\ \cline{2-5} 
                      & 1024   & 1           & 1           & 1           \\ \cline{2-5} 
                      & 2048   & 1           & 1           & 1           \\ \cline{2-5} 
                      & 4096   & 1           & 1           & .31         \\ \hline
\multirow{5}{*}{512}  & 256    & 0           & 0           & 0           \\ \cline{2-5} 
                      & 512    & 0           & 0           & 0           \\ \cline{2-5} 
                      & 1024   & 1           & .34         & 0           \\ \cline{2-5} 
                      & 2048   & 1           & 1           & 1           \\ \cline{2-5} 
                      & 4096   & 1           & 1           & .03         \\ \hline
\multirow{5}{*}{1024} & 256    & 0           & 0           & 0           \\ \cline{2-5} 
                      & 512    & 0           & 0           & 0           \\ \cline{2-5} 
                      & 1024   & 0           & 0           & 0           \\ \cline{2-5} 
                      & 2048   & 1           & .02         & 0           \\ \cline{2-5} 
                      & 4096   & 1           & .48         & 0           \\ \cline{2-5}
                      & 5120*   & 1           & 1           & 0           \\ \hline
\end{tabular}}
\caption{Fraction of $P$ recovered via gradient disaggregation for various participation rates. * indicates settings where the time limit for solving each column was increased to 60 minutes (vs 10 minutes). Generally, using more rounds facilitates more successful reconstruction; note that with larger number of users and rounds, success rate decreased due to exceeding the 10 minute time limit.}
\label{tab:extended_participation_rate}
\end{table}

\section{Extended Experiments on Constraint Granularity}

We provide extended data on gradient disaggregation against various parameter values of constraint granularity. Constraint granularity is how precise summary statistics capture user partipation frequency (see main paper for details). Unless stated, we enforced a time limit of 10 minutes for solving each column; we use a constraint granularity of 10. We show our extended results in Table \ref{tab:extended_granularity}.

\begin{table}[H]
\centering
\scalebox{.7}{
\begin{tabular}{|c|c|c|c|c|c|}
\hline
Users                 & Rounds & \multicolumn{4}{c|}{Constraint Granularity} \\ \hline
                      &        & 10  & 20   & 40   & \multicolumn{1}{l|}{80} \\ \hline
\multirow{5}{*}{128}  & 256    & 1   & 1    & .99  & .95                     \\ \cline{2-6} 
                      & 512    & 1   & 1    & 1    & 1                       \\ \cline{2-6} 
                      & 1024   & 1   & 1    & 1    & 1                       \\ \cline{2-6} 
                      & 2048   & 1   & 1    & 1    & 1                       \\ \cline{2-6} 
                      & 4096   & 1   & 1    & 1    & 1                       \\ \hline
\multirow{5}{*}{256}  & 256    & 0   & 0    & 0    & 0                       \\ \cline{2-6} 
                      & 512    & 1   & .94  & .27  & .02                     \\ \cline{2-6} 
                      & 1024   & 1   & 1    & .97  & .44                     \\ \cline{2-6} 
                      & 2048   & 1   & 1    & 1    & 1                       \\ \cline{2-6} 
                      & 4096   & 1   & 1    & 1    & .86                     \\ \hline
\multirow{5}{*}{512}  & 256    & 0   & 0    & 0    & 0                       \\ \cline{2-6} 
                      & 512    & 0   & 0    & 0    & 0                       \\ \cline{2-6} 
                      & 1024   & 1   & .5   & 0    & 0                       \\ \cline{2-6} 
                      & 2048   & 1   & 1    & .98  & .24                     \\ \cline{2-6} 
                      & 4096   & 1   & 1    & .38  & .035                    \\ \hline
\multirow{5}{*}{1024} & 256    & 0   & 0    & 0    & 0                       \\ \cline{2-6} 
                      & 512    & 0   & 0    & 0    & 0                       \\ \cline{2-6} 
                      & 1024   & 0   & 0    & 0    & 0                       \\ \cline{2-6} 
                      & 2048   & 1   & .23  & 0    & 0                       \\ \cline{2-6} 
                      & 4096   & 1   & .91  & 0    & 0                       \\ \hline
\end{tabular}}
\caption{Fraction of $P$ recovered via gradient disaggregation for various constraint granularities. }
\label{tab:extended_granularity}
\end{table}

\section{Extended Experiments on FedAvg Updates (Cifar10)}

We provide extended experiments on gradient disaggregation on FedAvg. We show our results in Table \ref{tab:exteneded_fedavg_1} and \ref{tab:exteneded_fedavg_2}.

\begin{table}[H]
\centering
\scalebox{.7}{
\begin{tabular}{|c|c|c|c|c|c|c|c|c|}
\hline
User Dataset Size                  & Batch Size & \multicolumn{7}{c|}{Local Epochs}                          \\ \hline
\multirow{5}{*}{384}               &            & 1   & 2   & 4   & \multicolumn{1}{l|}{8} & 16  & 32  & 64  \\ \cline{2-9} 
                                   & 8          & 1   & 1   & 1   & 1                      & 1   & 1   & 1   \\ \cline{2-9} 
                                   & 16         & 1   & 1   & 1   & 1                      & 1   & 1   & 1   \\ \cline{2-9} 
                                   & 32         & 1   & 1   & 1   & 1                      & 1   & 1   & 1   \\ \cline{2-9} 
                                   & 64         & 1   & 1   & 1   & 1                      & 1   & 1   & 1   \\ \hline
\multirow{4}{*}{384 (mom.=.9)}     & 8          & .96 & .88 & .78 & .97                    & .83 & .19 & .01 \\ \cline{2-9} 
                                   & 16         & 1   & 1   & 1   & 1                      & 1   & 1   & 1   \\ \cline{2-9} 
                                   & 32         & 1   & 1   & 1   & 1                      & 1   & 1   & 1   \\ \cline{2-9} 
                                   & 64         & 1   & 1   & 1   & 1                      & 1   & 1   & 1   \\ \hline
\multirow{4}{*}{384 (fraction=.9)} & 8          & .98 & 1   & 1   & 1                      & .99 & 1   & 1   \\ \cline{2-9} 
                                   & 16         & 1   & 1   & 1   & 1                      & 1   & 1   & 1   \\ \cline{2-9} 
                                   & 32         & 1   & 1   & 1   & 1                      & 1   & 1   & .99 \\ \cline{2-9} 
                                   & 64         & 1   & 1   & 1   & 1                      & 1   & 1   & 1   \\ \hline
\multirow{4}{*}{384 (fraction=.8)} & 8          & 1   & 1   & 1   & .99                    & .83 & .92 & 1   \\ \cline{2-9} 
                                   & 16         & .91 & .95 & .99 & 1                      & .97 & .84 & .95 \\ \cline{2-9} 
                                   & 32         & .99 & .99 & 1   & 1                      & 1   & 1   & 1   \\ \cline{2-9} 
                                   & 64         & 1   & 1   & 1   & 1                      & 1   & 1   & 1   \\ \hline
\end{tabular}}
\caption{Fraction of $P$ recovered on FedAvg updates using larger LeNet model (last hidden layer size=512), across various settings (mom.= SGD momentum, fraction=fraction of data sampled from the 384 examples to perform FedAvg over). Users=100, rounds=200, participation rate=.1, constraint granularity=10.}
\label{tab:exteneded_fedavg_1}
\end{table}

\begin{table}[H]
\centering
\scalebox{.7}{
\begin{tabular}{|c|c|c|c|c|c|c|c|}
\hline
Number of Users & Batch Size & \multicolumn{6}{c|}{Local Epochs}                  \\ \hline
                &            & 1    & 2 & 4    & \multicolumn{1}{l|}{8} & 16 & 32 \\ \hline
512             & 16         & 1    & 1 & 1    & .996                   & 1  & 1  \\ \hline
1024            & 16         & .998 & 1 & .999 & 1                      & 1  & 1  \\ \hline
\end{tabular}}
\caption{Fraction of $P$ recovered on FedAvg updates using larger LeNet model (last hidden layer size=512),  With more users. (rounds=200, participation rate=.1, constraint granularity=10). With many users, $P$ is still reconstructable.}
\label{tab:exteneded_fedavg_2}
\end{table}

\end{document}